# The Larmor frequency shift of a white matter magnetic microstructure model with multiple sources

Anders Dyhr Sandgaard[1], Noam Shemesh[2], Leif Østergaard[1], Valerij G. Kiselev[3] Sune Nørhøj Jespersen[1,4]

[1]Center of Functionally Integrative Neuroscience, Department of Clinical Medicine, Aarhus University, Denmark

[2]Champalimaud Research, Champalimaud Centre for the Unknown, Lisbon, Portugal,

[3]Division of Medical Physics, Department of Radiology, University Medical Center Freiburg, Freiburg, Germany

[4]Department of Physics and Astronomy, Aarhus University, Denmark

*Word Count: 12488 (whole document)*

Corresponding Author: Sune Nørhøj Jespersen.

Mail: sune@cfin.au.dk

## Keywords

Magnetic susceptibility, 2) Larmor frequency, 3) Magnetic microstructure, 4) Modelling, 5) Quantitative Susceptibility Mapping, 6) Lorentz cavity

## Abbreviations

**WM**: White Matter, **dMRI**: Diffusion MRI, **QSM**: Quantitative Susceptibility Mapping, **STI**: Susceptibility Tensor Imaging, **GLA**: Generalized Lorentzian Approach, **GLTA**: Generalized Tensor Approach, **MBP**: Myelin Basic Proteins, **PLP**: Proteo-Lipid Proteins, **MW**: Myelin Water, **fODF**: Fiber Orientation Distribution Function, **FA**: Fractional Anisotropy, **MD**: Mean Diffusivity, **FBI**: Fiber Ball Imaging, **GM**: Gray Matter, **SNR**: Signal-to-Noise Ratio, **RMSE**: Root-Mean-Squared Error, **DNR**: Diffusion Narrowing Regime (homogenous broadened spectral lineshape), **CSF**: Cerebral Spinal Fluid, **GR**: Tables of Gradshteyn and Ryzhik.

**Conflict of interest**

The authors declare no conflict of interest.

**Abstract**

Magnetic susceptibility imaging may provide valuable information about chemical composition and microstructural organization of tissue. However, its estimation from the MRI signal phase is particularly difficult as it is sensitive to magnetic tissue properties ranging from the molecular to macroscopic scale. The MRI Larmor frequency shift measured in white matter (WM) tissue depends on the myelinated axons and other magnetizable sources such as iron-filled ferritin. We have previously derived the Larmor frequency shift arising from a dense media of cylinders with scalar susceptibility and arbitrary orientation dispersion. Here we extend our model to include microscopic WM susceptibility anisotropy as well as spherical inclusions with scalar susceptibility to represent subcellular structures, biologically stored iron etc. We validate our analytical results with computer simulations and investigate the feasibility of estimating susceptibility using simple iterative linear least squares without regularization or preconditioning. This is done in a digital brain phantom synthesized from diffusion MRI (dMRI) measurements of an ex vivo mouse brain at ultra-high field.

## 1 | Introduction

The Larmor frequency of an MR-visible fluid in tissue is perturbed by microscopic variations in magnetic field induced by the sample. This microscopically varying magnetic field is determined by the tissue's magnetic susceptibility. In biological tissues, such microscopic perturbations are affected by the cellular and even subcellular tissue composition and configuration. Given the likely relevance of these tissue scales to early stages of disease[1–4], where subtle micro-architectural modulations may occur, it is highly desirable to accurately measure the magnetic tissue properties that drive the frequency perturbations. However, as the nominal MRI resolution on clinical scanners (typically mm) is orders of magnitudes greater than the microstructure (typically µm), such microscopic Larmor frequency perturbations become coarse-grained (fine details are smoothed) on the mesoscopic scale (typically 10-100µm) by both the measurement itself and by diffusion[5]. Nonetheless, a signature of the mesoscopically averaged magnetized microstructure remains embedded in the measured Larmor frequency shift[6–9].

Understanding how magnetic susceptibility of tissue gives rise to a measurable Larmor frequency shift has been an active field of research for decades[10–20]. In addition, efforts have focused on inverting this relationship to estimate local magnetic susceptibility in the brain and resulted in the common methods dubbed Quantitative Susceptibility Mapping (QSM)[21] and Susceptibility Tensor Imaging[22] (STI). Unfortunately, these inversion methods do not account for microscopic variations in the Larmor frequency shifts from susceptibility sources at microscopic distances from the reporting MR fluid. Failing to account for such microscopic features can ultimately lead to a substantial bias in susceptibility estimation[23] – especially in white matter (WM).

Incorporating such microscopic variations in the Larmor frequency shift into current QSM/STI models[24] requires explicit attention to the *microscopic magnetic anisotropy* of tissue in the sample. This includes *microscopic susceptibility anisotropy*, e.g., of the alkyl chain of the myelin sheath[25] and *microscopic structural anisotropy*, e.g., from the cylindrical geometry of WM axons (as can be seen in Figure 2). Both types of anisotropy give rise to an anisotropic Larmor frequency shift[8,25–30] that depends on the angle between the main magnetic field and directions of the axons. This has led to the development of several analytical models[11,13,14,30–33], e.g., describing the local microscopic Larmor frequency shift caused by *microscopic magnetic anisotropy* of WM. In particular, Wharton & Bowtell[13] and Sukstanskii & Yablonskiy[14] solved the magnetostatic equations to describe the microscopic Larmor frequency shift from a single infinitely long hollow cylinder accounting for both types of *microscopic magnetic anisotropy* of the myelin sheath. Sukstanskii & Yablonskiy[14] also considered each surrounding fluid compartment to have a scalar magnetic susceptibility different from the NMR fluid, e.g., to model iron-containing cells.

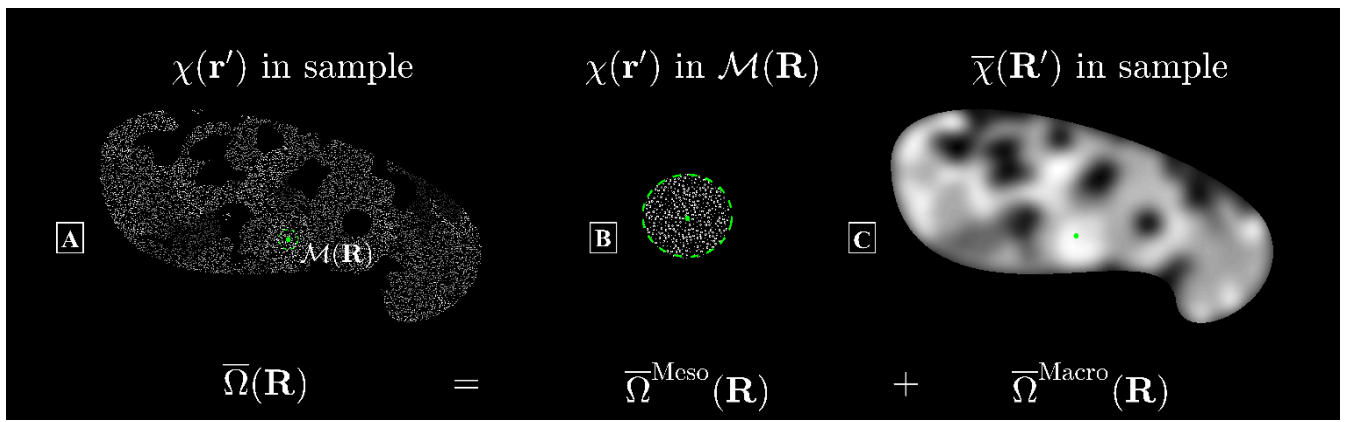

**Figure 1- Magnetic microstructure model:** Local magnetic susceptibility $\chi(r')$ from multiple regions with an associated microstructure. **A**: The coarse-grained Larmor frequency $\overline{\Omega}(\mathbf{R})$ is given by the microscopic Larmor

frequency $\Omega(r)$ felt by a spin at position $r$ averaged within a mesoscopic sphere $\mathcal{M}(R)$. $\overline{\Omega}(R)$ can be described by two contributions, the first of which is illustrated in **B**: the contribution from explicit microstructure within $\mathcal{M}(R)$, which defines the mesoscopic Larmor frequency $\overline{\Omega}^{\text{Meso}}(R)$. **C**: The second contribution depends on the Larmor frequency at $R$ induced by the mesoscopic averaged magnetic microstructure $\overline{\chi}(R')$. This defines the local macroscopic Larmor frequency $\overline{\Omega}^{\text{Macro}}(R)$. The image illustrates a mesoscopic spherical cavity in $\overline{\Omega}^{\text{Macro}}(R)$.

Due to the resolution of MRI measurement and the inherent diffusion of the MR-visible fluid, microscopic variations in the Larmor frequency must be coarse-grained on the mesoscopic scale in descriptions of the measured Larmor frequency shift. To account for this averaging, He & Yablonskiy[30] and later Yablonskiy & Sukstanskii[11,30], proposed a general framework dubbed the Generalized Lorentzian Approach (GLA) and Generalized Lorentzian Tensor Approach (GLTA), respectively. As illustrated in Figure 1, the approach followed the original concept of Lorentz[34]: Here, the main idea is to divide the whole sample (Figure 1A) into two regions around the point of interest: a *near* region (Figure 1B) and a *far* region (Figure 1C). The size of the near region should be larger than the characteristic size of the microstructure such that the frequency shift induced from sources in the far region can be described by a locally averaged magnetic susceptibility and the overall sample shape. This is the reason for calling the size of the near region *mesoscopic*[7,20,34], in comparison to the conventional molecular sized Lorentz cavity[7,12,20,34–36] used in the context of NMR to describe the frequency shift in isotropic liquids. The far region comprises the whole sample excluding a mesoscopic Lorentz cavity (the near region) around the point at which this coarse-grained Larmor frequency shift is considered. The frequency shift induced from sources within the near region must be calculated with account for the explicit magnetized microstructure and diffusion of the MR-visible fluid, as distances here are comparable with the characteristic length of the microstructure. Since the shape of this mesoscopic cavity can be chosen at our convenience[9], the main idea in the GL(T)A model is to identify a shape of the cavity (a so-called Lorentzian boundary) such that the mean frequency shift, which originates from the contained magnetized microstructure within it, can be neglected. By analogy, introducing a spherical molecular Lorentz cavity for isotropic liquids allows one to neglect an explicit calculation at the molecular scale[7,35] in isotropic liquids. On the mesoscopic scale, the shape of the cavity is less trivial as it relates to the total *magnetic anisotropy* of the microstructure. If the susceptibility is constant over the magnetized objects, the shape of the Lorentz cavity relates only to the *structural anisotropy* of the microstructure. If the microstructure is simple and its susceptibility constant, e.g., contains either uniformly magnetized parallel cylinders or spheres, the boundary is a coaxial cylinder or sphere, respectively. Hence, when this Lorentzian boundary of the mesoscopic cavity can be identified, such that an explicit calculation of the frequency shift inside the mesoscopic cavity can be avoided, only the averaged sources in the far region (Figure 1C) need to be considered, and one only has to account for the correction due to the shape of the subtracted mesoscopic cavity. In the GLTA model[11], this mesoscopic correction was defined by a so-called Lorentzian tensor, which contains information about the average susceptibility and microstructure inside the mesoscopic cavity (Figure 1B). As a model of WM microstructure, explicit expressions for the Lorentzian tensor were presented in the GLTA model for both solid parallel cylinders and spheres, assuming a low volume fraction and uniform susceptibility.

Explicit calculation of this Lorentzian tensor is generally difficult - especially for dense media including multiple types of inclusions - and identifying the appropriate shape of the mesoscopic cavity quickly becomes non-trivial. To overcome this, Ruh et al. & Kiselev[6,32] derived how the Lorentzian tensor for an arbitrary microstructure depends on its structural correlation function - as long as the susceptibility (scalar or tensor) remains uniform over the magnetized objects. The important difference here, compared to the GL(T)A model, is thinking of the microstructure in terms of the correlation function instead of the shape of the mesoscopic cavity. Ruh et al. & Kiselev[6,32] also reproduced the results for randomly positioned parallel solid cylinders with arbitrary volume fraction, and later, Sandgaard et al[9,24] extended the result to uniformly magnetized multi-layered cylinders with arbitrary orientations using the correlation function approach.

Current models of the Lorentzian tensor fail to capture salient WM features as they assume that WM axons occupy a low volume fraction, that WM axons are parallel, that microscopic magnetic susceptibility is uniform, etc. In reality, axon orientations are dispersed (at least by 20-30 degrees[37]), occupy a large fraction of the WM volume (around 30%), and possess non-uniform magnetic susceptibility[25,29]. Furthermore, experiments have demonstrated that biologically stored iron in WM, such as ferritin, affects the Larmor frequency shift[38–42]. All such realistic features should ideally be addressed in a model of the Larmor frequency shift in WM.

In this study, we derive the Lorentzian tensor for non-uniform microscopic magnetic susceptibility and present new analytical results for the mesoscopically averaged (coarse-grained) Larmor frequency shift. Our biophysical model of WM consists of a dense media of orientationally dispersed, multi-layered cylinders with microscopic susceptibility anisotropy to model axons, and spherical inclusions in all major water compartments to model subcellular structures, biologically stored iron etc. We describe how this complicated microstructure gives rise to a set of effective susceptibility parameters from all the susceptibility sources, and 6 structural parameters that describe the axonal orientation dispersion, which can be determined independently from diffusion MRI (dMRI).

## 2 | Theory

In this section, we first outline the system of consideration. We then revisit the theoretical framework for the mesoscopically averaged Larmor frequency shift $\bar{\Omega}(\mathbf{R})$. Third, we present analytical results for a biophysical model of WM magnetic microstructure incorporating both orientationally dispersed WM axons with microscopic susceptibility anisotropy and spherical inclusions representing e.g., biologically stored iron. Last, our main analytical result for the model-specific mesoscopic Larmor frequency shift $\bar{\Omega}(\mathbf{R})$ is presented, while a complete derivation can be found in supplementary material.

**System of consideration**

The macroscopic sample of volume $V$ is described as a porous medium of impermeable microscopic magnetizable inclusions immersed in an MR-visible fluid. The inclusions are magnetized by the external field $\mathbf{B}_0 = B_0 \hat{\mathbf{B}}$, where $\hat{\mathbf{B}}$ is a unit vector. All inclusions are assumed to be weakly dia- or paramagnetic, and characterized for each inclusion by a magnetic susceptibility tensor $\chi^Q(\mathbf{r})$ ( $|\chi| \ll 1$, and is given relative to the susceptibility of the MR fluid with susceptibility $\chi^W$ and volume fraction $\zeta^W$ ).

**Mesoscopic Larmor frequency shift $\bar{\Omega}(\mathbf{R})$:**

The Larmor frequency $\bar{\Omega}(\mathbf{R})$ describes the average microscopic frequency shifts felt by the MR-visible fluid inside a mesoscopic region $\mathcal{M}$ at position $\mathbf{R}$ (cf. Figure 1). When $\bar{\Omega}(\mathbf{R})$ does not vary across the sampling point spread function, it provides an approximation to the MRI measured Larmor frequency shift, i.e., $\overline{\Omega}_{\text{MRI}}(\mathbf{R}) = \overline{\Omega}(\mathbf{R})$ inside a voxel at position $\mathbf{R}$ [9,24,32]. This mesoscopic Larmor frequency shift can be described by a mesoscopic contribution $\overline{\Omega}^{\text{Meso}}(\mathbf{R})$, which is the mean induced frequency shift in $\mathcal{M}$ from the actual distribution of sources within $\mathcal{M}$, a macroscopic contribution $\overline{\Omega}^{\text{Macro}}(\mathbf{R})$ describing the mean induced Larmor frequency shift in $\mathcal{M}$ from distant sources outside $\mathcal{M}$, and an additional contribution $\overline{\Omega}^{\text{W}}(\mathbf{R})$ describing the mean induced Larmor frequency shift in $\mathcal{M}$ from the whole sample with susceptibility $\chi^{\text{W}}$

$$\overline{\Omega}(\mathbf{R}) = \overline{\Omega}^{\text{Meso}}(\mathbf{R}) + \overline{\Omega}^{\text{Macro}}(\mathbf{R}) + \overline{\Omega}^{\text{W}}(\mathbf{R}). \tag{1}$$

This decomposition is visualized in Figure 1. In general, the shape of $\mathcal{M}$ can always be chosen for our convenience. Here we choose $\mathcal{M}$ to be spherical, as it represents the best compromise when working with a general magnetic microstructure. While the shape of $\mathcal{M}$ may affect the values of $\overline{\Omega}^{\text{Meso}}(\mathbf{R})$ and $\overline{\Omega}^{\text{Meso}}(\mathbf{R})$ individually, the sum $\overline{\Omega}^{\text{Meso}}(\mathbf{R}) + \overline{\Omega}^{\text{Macro}}(\mathbf{R})$ remains unchanged. Each contribution will be described in turn.

*Mesoscopic contribution $\overline{\Omega}^{\text{Meso}}(\mathbf{R})$*

The microscopic Larmor frequency shift sensed by the MR fluid depends in general on time due to water diffusion. This in turn means that the mesoscopic contribution $\overline{\Omega}^{\text{Meso}}(\mathbf{R})$ to $\overline{\Omega}(\mathbf{R})$, may depend on time. However, as described in a previous study[9]: considering long times in comparison to the diffusion time $\tau_c = l_c^2 / D$ of the microstructure, where $l_c$ is the characteristic length of the microstructure and $D$ the diffusivity, the mesoscopic contribution becomes time independent. This regime is the so-called diffusion-narrowing regime (DNR)[43]. Here $\overline{\Omega}^{\text{Meso}}(\mathbf{R})$ is described by a Lorentzian tensor $\mathbf{L}$ of the magnetic microstructure inside a mesoscopic sphere $\mathcal{M}$ [6,7,30,34] surrounding $\mathbf{R}$ (see previous work[9] on the freedom of choosing the shape of $\mathcal{M}$)

$$\overline{\Omega}^{\text{Meso}}(\mathbf{R}) = \gamma B_0 \hat{\mathbf{B}}^{\text{T}} \mathbf{L} \hat{\mathbf{B}}, \tag{2}$$

where

$$\mathbf{L} = \frac{1}{\zeta^W} \int \frac{d\mathbf{k}}{(2\pi)^3} \Upsilon(\mathbf{k}) \Gamma(\mathbf{k}). \qquad (3)$$

Here $\Upsilon(\mathbf{k})$ is the elementary dipole kernel in Fourier space including a molecular Lorentz correction[35] and $\zeta^W$ the total water volume fraction outside all inclusion types. Eq. (3) depends in turn on the tensor-valued magneto-structural cross-correlation function[9] $\Gamma(\mathbf{k})$

$$\Gamma(\mathbf{k}) = \frac{v^W(\mathbf{k})\chi(-\mathbf{k})}{|\mathcal{M}|}, \quad k > 0, \qquad (4)$$

and zero for $k = 0$. For notational simplicity, we leave it understood implicitly that the cross-correlations in Eqs. (2)-(4) are considered in the vicinity of $\mathbf{R}$. The size $|\mathcal{M}|$ denotes the volume of the mesoscopic sphere needed[9] to average the magnetic microstructure near $\mathbf{R}$, and to allow the exterior to be described by an averaged (bulk) magnetic microstructure[9] $\bar{\chi}(\mathbf{R})$. Here $v^W(\mathbf{k})$ is the total indicator function of the reporting water surrounding the microscopic inclusions while $\chi(\mathbf{k})$ denotes the microscopic susceptibility tensor of the inclusions in k-space. Our definition of the Lorentzian tensor $\mathbf{L}$ differs from the GLTA model: here an isotropic media gives $\mathbf{L} = \mathbf{0}$, while the GLTA model predicts $\mathbf{L}_{\text{GLTA}} = \mathbf{I}/3$. This does not mean the two models disagree, as it is merely a rearrangement of terms in $\bar{\Omega}^{\text{Meso}}(\mathbf{R}) + \bar{\Omega}^{\text{Macro}}(\mathbf{R})$ such that the sum remains unchanged.

If the magnetic susceptibility is constant inside the inclusions (scalar or tensor), $\mathbf{L} = -\mathbf{N}\chi$ is described by a demagnetization tensor $\mathbf{N}$

$$\mathbf{N} = -\frac{1}{\zeta^W} \int \frac{d\mathbf{k}}{(2\pi)^3} \Upsilon(\mathbf{k}) \Gamma(\mathbf{k}), \qquad (5)$$

which depends only on the structure-structure correlation function

$$\Gamma(\mathbf{k}) = \frac{v^W(\mathbf{k}) v(-\mathbf{k})}{|\mathcal{M}|}, \quad k > 0. \qquad (6)$$

In general, the MR fluid is weighted by differences in transverse relaxation, but if all the fluid contributes equally to the frequency shift in $\mathcal{M}$, the fluid indicator function $v^W(\mathbf{r}) = 1 - v(\mathbf{r})$ relates to the structural indicator function $v(\mathbf{r})$ of all inclusions. By choosing $\mathcal{M}$ as a sphere, there is no need to consider the 1 in $v^W$ in Eq. (4) since it does not contribute when substituted in Eq. (3) (corresponding to the Lorentz-corrected field inside a homogenously magnetized sphere[35]). This means that the mean magnetic field in $v^W(r)$ is equal and opposite in sign to the field inside $v(r)$. The correlation tensor $\Gamma(k)$ extends the previous description of $\mathbf{L}(\mathbf{R})$ to include non-uniform susceptibility, while previous work described uniformly magnetized inclusions reducing $\Gamma(k)$ to the structural correlation function $\Gamma(k)$.

*Macroscopic contribution $\overline{\Omega}^{Macro}(\mathbf{R})$*

The mesoscopically averaged Larmor frequency shift $\overline{\Omega}^{Macro}(\mathbf{R})$ inside $\mathcal{M}$ caused by sources outside of $\mathcal{M}$ does not depend on explicit microscopic details of $v(r)$ and $\chi(r)$ - only averaged properties[9] like the bulk magnetic susceptibility $\overline{\chi}(\mathbf{R})$ matter:

$$\overline{\Omega}^{Macro}(\mathbf{R}) = \gamma B_0 \hat{\mathbf{B}}^T \int \Upsilon(\mathbf{R}-\mathbf{R}')\overline{\chi}(\mathbf{R}')\hat{\mathbf{B}}, \ (\mathcal{M} \text{ is spherical}). \tag{7}$$

Since $\mathcal{M}$ was chosen as a sphere, integration can be performed over the whole sample volume, as integrating $\overline{\chi}(\mathbf{R})$ inside $\mathcal{M}$ yields zero due to the Lorentz-sphere correction in $\Upsilon$. If the shape of $\mathcal{M}$ is different, e.g., a cylinder, integration over $\mathcal{M}$ must be subtracted in Eq. (7). Subtracting a cylindrical cavity in $\overline{\Omega}^{Macro}(\mathbf{R})$ should therefore not be forgotten, when including the mean microscopic field described by previous single hollow cylinder models[13]. In MRI, the Larmor frequency shift is sampled at discrete positions $\mathbf{R}'$. In this case, the macroscopic contribution $\overline{\Omega}^{Macro}(\mathbf{R})$ can be approximated on the level of the sampling resolution by a discrete convolution between a voxel-averaged dipole field[9] $\overline{\Upsilon}(\mathbf{R})$ and the averaged magnetic susceptibility $\overline{\chi}(\mathbf{R}')$ across the sample

$$\overline{\Omega}^{Macro}(\mathbf{R}) \approx \gamma B_0 \hat{\mathbf{B}}^T \sum_{\mathbf{R}'} \overline{\Upsilon}(\mathbf{R}-\mathbf{R}')\overline{\chi}(\mathbf{R}')\hat{\mathbf{B}}. \tag{8}$$

Here the whole sample has been partitioned into voxels at discrete positions $\mathbf{R}'$ with size described by the sampling resolution. The sample shape is thus represented implicitly by the sum. Notice that Eq. (8) corresponds to the frequency shift considered in STI and QSM, albeit they typically use the elementary dipole field $\Upsilon(\mathbf{R})$ and not the voxel-averaged $\overline{\Upsilon}(\mathbf{R})$, which can lead to a bias in parameter estimation[9].

*Frequency shift $\overline{\Omega}^W(\mathbf{R})$ from MR fluid*

The last frequency shift from the MR fluid depends on its susceptibility $\chi^W$ and the sample shape

$$\overline{\Omega}^W(\mathbf{R}) = \chi^W \gamma B_0 \hat{\mathbf{B}}^T \mathbf{N}^W(\mathbf{R}) \hat{\mathbf{B}}. \tag{9}$$

Here $\mathbf{N}^W(\mathbf{R})$ defines the sample-specific Lorentz-corrected point-demagnetization tensor [44,45]

$$\mathbf{N}^W(\mathbf{R}) = \int d\mathbf{R}' \, \Upsilon(\mathbf{R}-\mathbf{R}'), \tag{10}$$

where the integration is performed over the sample.

**A white matter magnetic microstructure model**

To develop robust biophysical models of WM magnetic microstructure that describes the MRI measured Larmor frequency shift, it is important to identify the main features affecting the microscopic Larmor frequency sensed by water molecules. As describes previously[24], there are two main types of *microscopic magnetic anisotropy* affecting the microscopic Larmor frequency shift: *microscopic structural anisotropy* and *microscopic susceptibility anisotropy.* This means that a model should not only embrace microstructural features of tissue, but also susceptibility features due to chemical composition.

In WM, most axons in the central nervous system are insulated by a myelin sheath spiraling around the axon[46]. Figure 2 gives an overview of the construction. The myelin sheath consists of alternating layers of protein rich water and lipids. Each lipid sheath is constructed mainly from phospholipids, glycolipids, and cholesterol. Their hydrophobic tails combine to form a lipid bilayer such that their hydrophilic heads are in direct contact to the water-filled protein layers, hampering the mobility of water molecules near the interface of the bilayers[47]. The inner protein layers contain amphipathic myelin basic proteins (MBP) which stabilize the membrane by interacting with the charged heads of the lipids. The outer protein layers are connected to the neighboring membranes via phospholipid protein chains (PLP) and are openly connected to the extra-cellular space. PLP form channels for transporting ions or small molecules across the membrane. The water filled protein layers are extremely thin (around 2-4 nm thick) which is around 10 times the size of a water molecule (around 0.3 nm). These protein rich water compartments are collectively denoted as myelin water (MW). The myelin membranes are produced by externally placed oligodendrocytes interposed between different myelinated axons and can be responsible for myelination of up to 40 myelin segments of different adjoining axons. An important component in myelin production is iron, and oligodendrocytes are the most predominant iron-containing cell in WM[48]. Transferrin is responsible for transporting iron atoms to the oligodendrocytes[48]. If not immediately used in myelin production, the iron atoms are stored in ferritin complexes. A full ferritin complex can store up to 4500 iron atoms[49]. The highest WM iron concentrations are found by iron staining to be in the superficial WM, while deep white matter, such as the corpus callosum and optic radiation, are devoid of staining[50].

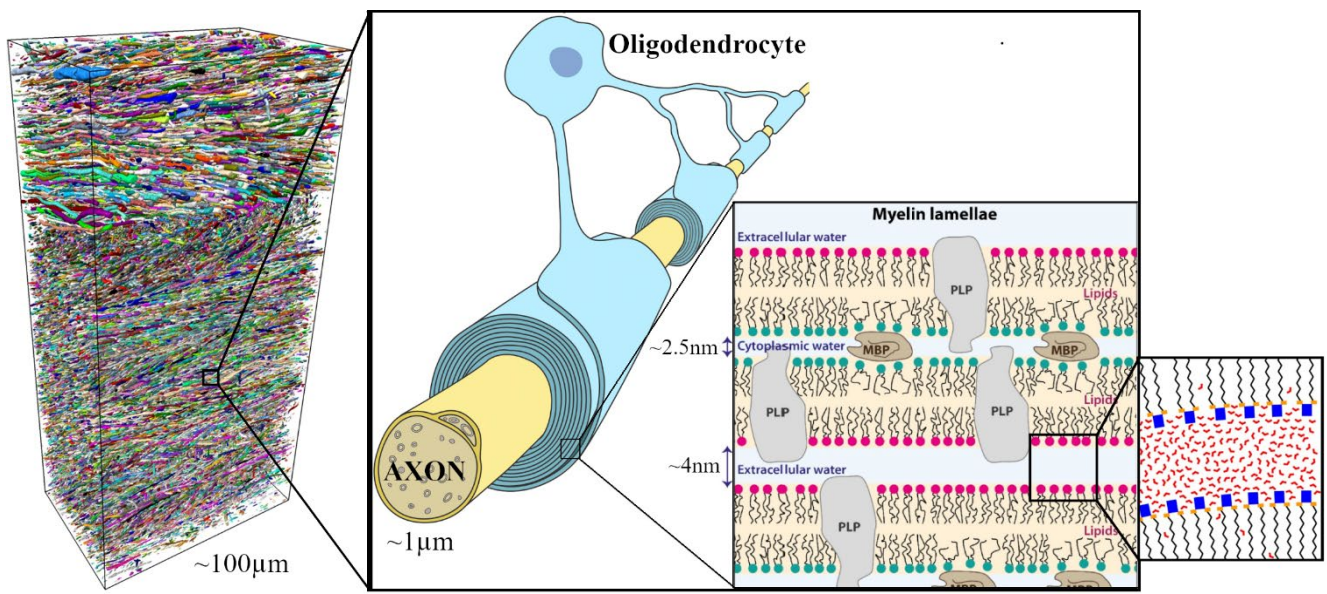

**Figure 2 – Depiction of the myelinated axons in white matter:** To the left, 3D segmentation of myelinated axons in corpus callosum and cingulum of a rat is shown[51]. As shown to the right, a myelin sheath is wrapped around the axons, produced by a neighboring oligodendrocyte. The myelin lamellae are comprised of alternating layers of lipid bilayers (yellow background) and protein-rich water layers (blue background). The hydrophobic tails of the lipids attract to form the lipid bilayer, leaving the hydrophilic heads (green and red circles) in direct connection with the waterfilled protein layers (cytoplasmic and extracellular water). The myelin basic proteins (MBP) stabilize the lamellae, while the phospholipid proteins (PLP) form channels to allow transportation of ions or small molecules to pass through. As shown to the right, water molecules in protein rich myelin water form a hydration later near the hydrophilic heads of the lipids.

The illustration is adapted and reprinted from[51], [52] and [53] with permission from Elsevier, Creative Commons and Wiley materials (Copyright © 2013 Wiley Periodicals, Inc.), respectively. The image contains illustrations from [54] with permission from PNAS (exempt from Creative Commons license).

The main goal of this study is to generalize existing models[11,13,14,32,36,55] of the Larmor frequency shift in WM described in the introduction. Our model of WM magnetic microstructure embraces the most contributing microscopic susceptibility sources, including their structural anisotropy, expected to affect the measured Larmor frequency shift the most. Our new magnetic microstructure model of WM (as seen in Figure 3) consists of the following two main groups of sources assumed to be impermeable for water molecules:

**Infinitely long cylinders** randomly positioned with an arbitrary size distribution independent of their orientations. Each cylinder consists of multiple concentric layers to model the bilayers consisting of lipids and PLP channels etc. Each layer has an axially symmetric microscopic susceptibility tensor $\boldsymbol{\chi}^C$, which models the radially positioned lipid bilayers[30] such that the magnetic response is $\chi_{\parallel}$ parallel to each lipid and $\chi_{\perp}$ perpendicular [25]. Their positions are described by a microscopic indicator function $v^C(\boldsymbol{r})$ (which is 1 inside inclusions and 0 otherwise), and their total volume fraction by $\zeta^C$. This extends our previous proposed model for solid and multilayered cylinders with arbitrary orientation dispersion[9], and goes beyond previous descriptions of WM magnetic tissue heterogeneity. The magnetic susceptibility $\boldsymbol{\chi}^C$ of each layer-forming lipid inclusion of the axons pointing along the radial direction $\hat{\mathbf{u}}$ with respect to the cylinder becomes (cf. Figure 3A)

$$\boldsymbol{\chi}^C = \chi_{\perp}\left(\mathbf{I} - \hat{\mathbf{u}}\hat{\mathbf{u}}^T\right) + \chi_{\parallel}\hat{\mathbf{u}}\hat{\mathbf{u}}^T = \left(\chi^C - \frac{1}{3}\Delta\chi\right)\mathbf{I} + \Delta\chi\hat{\mathbf{u}}\hat{\mathbf{u}}^T \quad \text{(myelin lipid)}. \tag{11}$$

Here

$$\chi^C \equiv \frac{\chi^{\parallel} + 2\chi^{\perp}}{3} = \frac{1}{3}\mathrm{Tr}(\boldsymbol{\chi}), \text{ and } \Delta\chi = \chi_{\parallel} - \chi_{\perp} \tag{12}$$

defines one third the trace of $\boldsymbol{\chi}^C$ and the microscopic susceptibility anisotropy, respectively.

**Spherical inclusions** randomly positioned in all major water compartments (cf. Figure 3B). Each water compartment contains its own population of spherical inclusions. Each population may reflect multiple spherical sources with different susceptibility. Here we focus on a single population of spheres in each water compartment for simplicity, but this can naturally be extended to multiple different sources. Hence, each compartmental susceptibility represent a weighted average across all susceptibility sources residing in a given compartment. Spherical inclusions (such as biologically stored iron, e.g., iron filled ferritin molecules) are assumed to reside in the extra-axonal space with magnetic susceptibility denoted $\chi^E$ with volume fraction $\zeta^E$. The distribution of extra-axonal spheres is described by the indicator function $v^E(r)$. The distribution of spherical inclusions in the intra-axonal space is described by $v^A(r)$ with magnetic susceptibility $\chi^A$ and volume fraction $\zeta^A$. The MW layers are assumed to include spherical inclusions such as diamagnetic MBP and other potential sources, and here the susceptibility is denoted $\chi^M$, the volume fraction $\zeta^M$ and the indicator function $v^M(r)$. All the different spherical inclusions in the water compartments have a total indicator function given by the sum $v^S(r) = v^A(r) + v^M(r) + v^E(r)$ and susceptibility $\chi^S(r) = \chi^A v^A(r) + \chi^M v^M(r) + \chi^E v^E(r)$. In our model spheres can have an arbitrary volume fraction, but in practice they are assumed to occupy a low volume fraction in WM, i.e., $\zeta^S = \zeta^A + \zeta^M + \zeta^E \ll 1$, which is justified by histological findings [56].

*Water volume fraction $\zeta^W$ and indicator function $v^W$*

The total volume fraction of all inclusions is $\zeta = \zeta^C + \zeta^M + \zeta^E + \zeta^A$. The indicator function $v^W(r)$ of the reporting water, Eq. (3), is in general weighted by the signal attenuation caused by transverse relaxation, which sets it apart from the purely geometrical indicator functions of the inclusions. If all the water contributes evenly, the water indicator function becomes $v^W(r) = 1 - v^C(r) - v^S(r)$ with $\zeta^W = 1 - \zeta$ describing the total water fraction. If MW is fully relaxed, the water indicator function becomes $v^W(r) = 1 - v^C(r) - v^S(r) - v^{MW}(r)$ with water volume fraction $\zeta^W = 1 - \zeta^C - \zeta^S - \zeta^{MW}$, where $\zeta^{MW}$ denotes the total water volume fraction of the MW compartment. We assume water to be moving around freely in all water compartments. While the assumption of freely moving water may be violated in MW[53,57,58], the main focus of this study is to consider the case when MW is fully relaxed, as is the case in DNR [43] due to the rapid signal decay of MW[59,60].

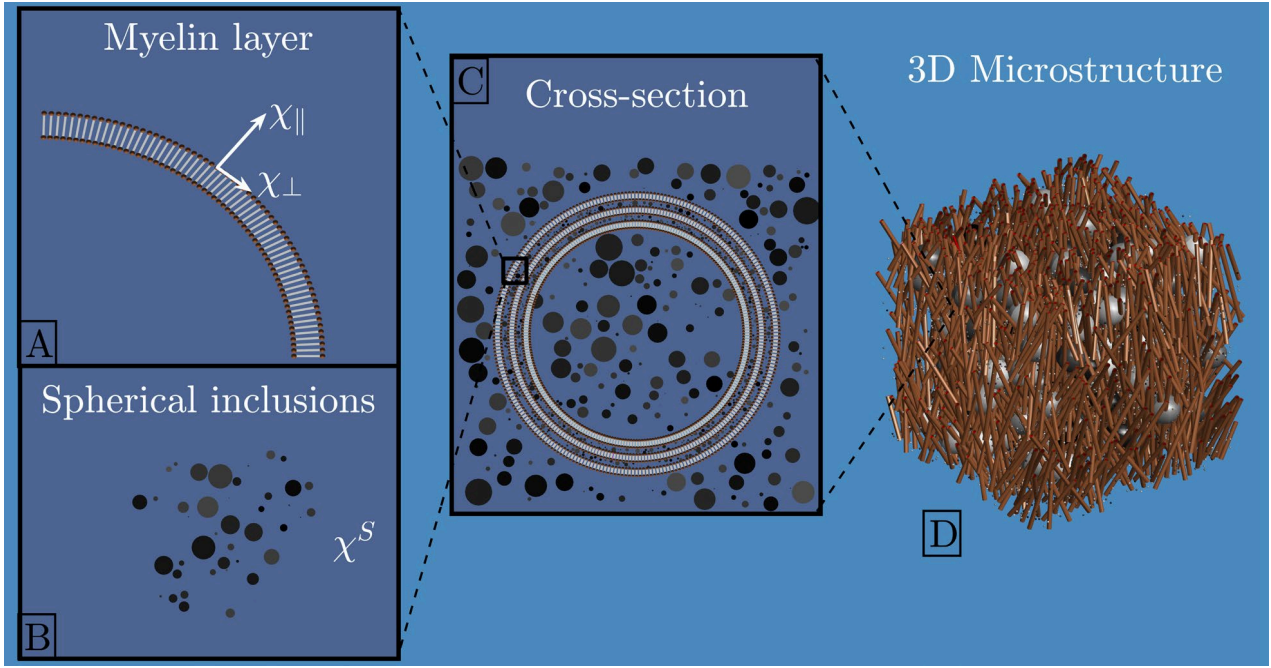

**Figure 3 – Magnetic microstructure model from different perspectives: A** shows the myelin shell assumed to consist of lipid chains and proteolipid proteins. The chains exhibit microscopic susceptibility anisotropy $\Delta\chi = \chi_\parallel - \chi_\perp$. **B** depicts the different spherical inclusions which can be found in all water layers and can have different sizes and susceptibilities (here shown by varying color and size). **C** conceptualizes the magnetic microstructure of a single axon with susceptibility sources shown in **A** and **B**. Spherical inclusions with isotropic susceptibility $\chi^E$ are randomly positioned outside cylinders and could mimic iron complexes outside axons. Within the bi-layers, myelin basic proteins (large spheres) or point-like particles are shown. They possess an isotropic susceptibility $\chi^M$. Intra-axonal point-like spheres with isotropic susceptibility $\chi^A$ are also included. Their total susceptibility is $\chi^S = \chi^A + \chi^M + \chi^E$ and assumed to occupy a low volume fraction in WM. **D** shows the entire 3D magnetic microstructure to demonstrate orientation dispersion and other spherical inclusions e.g., neuroglia. All the proportions are exaggerated for illustrative purposes.

**Lorentzian tensor of WM magnetic microstructure model**

Since two different susceptibility sources $\chi^C$ and $\chi^S$ are considered here, it is convenient to separate the full Lorentzian tensor $\mathbf{L}$ from both sources into two contributions, cf. Eq. (3), since $\mathbf{L}$ is linear with respect to susceptibility. This means that $\mathbf{L}$ can be described by two Lorentzian tensors $\mathbf{L}^C$ and $\mathbf{L}^S$, where $\mathbf{L}^C$ corresponds to the mesoscopic contribution induced by $\chi^C$ and $\mathbf{L}^S$ the mesoscopic contribution induced by the spherical susceptibility $\chi^S$:

$$\mathbf{L} = \mathbf{L}^C + \mathbf{L}^S. \qquad (13)$$

Here the Lorentzian tensor $\mathbf{L}^C$ is

$$\mathbf{L}^C = -\frac{1}{\zeta^W}\int \frac{d\mathbf{k}}{(2\pi)^3}\Upsilon(\mathbf{k})\left(v^{MW}(\mathbf{k}) + v^S(\mathbf{k}) + v^C(\mathbf{k})\right)\chi^C(-\mathbf{k}), \text{(Cylinders are sources)} \qquad (14)$$

The Lorentzian tensor $\mathbf{L}^C$ describes the measured Larmor frequency shift inside $\mathcal{M}$ from the cylinders with non-uniform susceptibility $\chi^C$ as described by Eq. (11).

The Lorentzian tensor $\mathbf{L}^S$ is

$$\mathbf{L}^S = -\frac{1}{\zeta^W}\int \frac{d\mathbf{k}}{(2\pi)^3}\Upsilon(\mathbf{k})\left(v^{MW}(\mathbf{k}) + v^S(\mathbf{k}) + v^C(\mathbf{k})\right)\chi^S(-\mathbf{k}), \text{(Spheres are sources)} \qquad (15)$$

Here the Lorentzian tensor $\mathbf{L}^S$ describes the measured Larmor frequency shifts inside $\mathcal{M}$ from all the spheres with uniform susceptibility $\chi^S(\mathbf{k}) = \chi^A v^A(\mathbf{k}) + \chi^M v^M(\mathbf{k}) + \chi^E v^E(\mathbf{k})$. Notice that $\mathbf{L}^S$ depends only on structural correlation functions taking scalar values, which mean it consists of a sum of demagnetization tensors $\mathbf{N}$, Eq. (5). In contrast, $\mathbf{L}^C$ depends on tensor-valued magneto-structural correlation functions. As shown in supporting material S1, these correlation functions are constructed from the indicator functions and magnetic susceptibilities of cylinders and spheres in k-space. We present our main results below, while the complete (but rather lengthy) derivation can be found in the supporting material along with numerical simulations for validation of the analytical results.

*Cylindrical inclusions*

Consider the indicator function for a single multi-layered cylinder $v(\mathbf{k})$ in k-space (see supporting material S1) positioned at $\mathbf{u}$ and pointing along $\hat{\mathbf{n}}$. Letting $r_q$, $R_q$ denote the inner and outer radii for the q'th cylinder layer, respectively, $v(\mathbf{k})$ becomes

$$v(\boldsymbol{k}) = e^{i\boldsymbol{k}\cdot\boldsymbol{u}} \frac{4\pi^2}{k} \sum_q \left( R_q J_1(R_q k) - r_q J_1(r_q k) \right) \delta(\boldsymbol{k}\cdot\hat{\boldsymbol{n}}). \text{ (multi-layered cylinder)}. \qquad (16)$$

The non-uniform susceptibility tensor $\boldsymbol{\chi}^C(\boldsymbol{k})$ for a single multi-layered cylinder in k-space are found to be

$$\boldsymbol{\chi}^C(\boldsymbol{k}) = 4\pi^2 \delta(\boldsymbol{k}\cdot\hat{\boldsymbol{n}}) e^{i\boldsymbol{k}\cdot\boldsymbol{u}} \left\{ \boldsymbol{\chi}_0 \sum_q \left( \xi_0(kR_q) - \xi_0(kr_q) \right) \right. \\ \left. - \text{Re}\{\boldsymbol{\chi}_2 e^{2i\psi}\} \sum_q \left( \xi_2(kR_q) - \xi_2(kr_q) \right) \right\}, \text{ (Multi-layered cylinder)}. \qquad (17)$$

Here

$$\boldsymbol{\chi}_0 = \chi^C \mathbf{I} + \frac{1}{2}\Delta\chi\left(\frac{1}{3}\mathbf{I} - \hat{\boldsymbol{n}}\hat{\boldsymbol{n}}^T\right) \text{ and } \boldsymbol{\chi}_2 = \frac{\Delta\chi}{2}\left(\hat{\boldsymbol{u}}\hat{\boldsymbol{u}}^T - \hat{\boldsymbol{v}}\hat{\boldsymbol{v}}^T - i\left(\hat{\boldsymbol{u}}\hat{\boldsymbol{v}}^T + \hat{\boldsymbol{v}}\hat{\boldsymbol{u}}^T\right)\right) \qquad (18)$$

corresponds to the only non-zero coefficient matrices of the Fourier series of $\boldsymbol{\chi}^C(\boldsymbol{k})$, and $\psi$ denotes the azimuthal angle in the local coordinate system coaxial to the cylinder orientation $\hat{\boldsymbol{n}}$. If $\Delta\chi = 0$, $\boldsymbol{\chi}^C(\boldsymbol{k}) = \chi^C v(\boldsymbol{k})$ which has already been considered in previous studies[9]. The magnetic susceptibility of multiple cylinders is then simply a sum over all cylinders. The bulk susceptibility $\bar{\boldsymbol{\chi}}^C$ which is used in the macroscopic contribution $\bar{\Omega}^{\text{Macro}}$, Eq. (8), is

$$\bar{\boldsymbol{\chi}}^C = \bar{\chi}^C \mathbf{I} + \frac{\Delta\bar{\chi}}{2}\left(\frac{1}{3}\mathbf{I} - \mathbf{T}\right), \qquad (19)$$

which follows simply from summing over all cylinders labelled by the index q $\bar{\boldsymbol{\chi}}^C = \sum_q \boldsymbol{\chi}_{0q}(k=0)/|\mathcal{M}|$. The parameters $\bar{\chi}^C = \zeta^C \chi^C$ and $\Delta\bar{\chi} = \zeta^C \Delta\chi$ define the orientationally invariant bulk magnetic susceptibilities. Here $\mathbf{T} = \sum_j \hat{\boldsymbol{n}}\hat{\boldsymbol{n}}^T / N$ is the scatter matrix of the fiber orientation distribution (fODF)[9,61], and is a rank-2 tensor describing the orientation dispersion of the fODF. Hence, the effective susceptibility anisotropy $\Delta\bar{\chi}$ on the mesoscopic scale depends on the fODF and the amount of orientation dispersion.

*Spherical inclusions*

The k-space indicator function of a generic sphere $v(\boldsymbol{k})$ with radius $R$ and positioned at $\boldsymbol{u}$ is (see supporting material S2)

$$v(\mathbf{k}) = e^{i\mathbf{k}\cdot\mathbf{u}} 4\pi R^3 \frac{j_1(kR)}{kR}, \text{ (generic sphere)}. \tag{20}$$

Hence, to determine the model-predicted Larmor frequency shift $\bar{\Omega}$, all we need to do is compute the two tensors $\mathbf{L}^C$ and $\mathbf{L}^S$ using the derived indicator functions. The integrals describing the Lorentzian tensors are included as supporting material. The next section presents the results for $\mathbf{L}^C$ and $\mathbf{L}^S$.

**Mesoscopic Larmor frequency shift of WM magnetic microstructure model**

In this section, we provide the final analytical results for the mesoscopic Larmor frequency shift $\bar{\Omega}$ from all the different inclusions. The main results are presented separately for the contribution induced by cylinders with susceptibility $\chi^C$ ($\mathbf{L}^C$ cf. Eq. (14)), and the contribution generated by spheres with $\chi^S$ ($\mathbf{L}^S$ cf. Eq. (15)).

*Larmor frequency shift from cylindrical inclusions*

For a distribution of multi-layered cylinders, we find for the Lorentzian tensor $\mathbf{L}^C$ (cf. Eq. (14))

$$\mathbf{L}^C = -\bar{\chi}^C \frac{1}{2}\left(\mathbf{T} - \frac{1}{3}\mathbf{I}\right) + \Delta\bar{\chi}\frac{1}{12}\left((1-\lambda)\mathbf{T} + \left(\lambda + \frac{1}{3}\right)\mathbf{I}\right). \tag{21}$$

Here $\lambda$ is a geometric factor (see Eqs. (S21) and (S65) in supporting material) depending on the layer geometry of the axons and their internal water fractions. Equation (21) is valid whether MW is relaxed or not, but its saturation state is reflected in the value of $\lambda$. For example, in the case of fully relaxed MW and axons of similar size with many thin layers[14],

$$\lambda = -6\frac{\zeta^{AW}}{\zeta^C \zeta^W}\frac{d}{d+d^W}\ln(g) > 0, \text{ (Equally sized cylinder layers)}, \tag{22}$$

Where $\zeta^{AW}$ is the intra-axonal water volume fraction, $g = r_1/R_N$ is the ratio between the radii of the inner $r_1$ and outer cylinder layer $R_N$, $d$ is the width of the lipid layers and $d^W$ the width of the MW layers. This means that cell bodies, neurofilaments, microtubules etc. which can be found in the intra-axonal space, decrease $\lambda$, due to the sensitivity to $\zeta^{\text{AW}}$. This is because the mesoscopic frequency shift, cf. Eq. (3), is considered here over all the water inside $\mathcal{M}$, which is normalized by the water volume fraction $\zeta^{\text{W}}$. Hence, it is obvious that the less water is present in the intra-axonal space, the less it contributes to the mesoscopic frequency shift $\bar{\Omega}(\mathbf{R})$. If the cylinders have different size, Eq. (S65) in supporting material must be averaged according to the size distribution[62].

The scatter matrix $\hat{\mathbf{B}}^{\text{T}}\mathbf{T}\hat{\mathbf{B}}$ along the measured direction $\hat{\mathbf{B}}$, cf. Eq. (2), can be estimated using dMRI via the relation[24,63]

$$\mathbf{T} - \frac{1}{3}\mathbf{I} = \frac{2}{15}\sum_{m=-2}^{2} p_{2m}\mathcal{Y}_{2m}, \qquad (23)$$

Here $\mathcal{Y}_{2m}$ defines the $l=2$ symmetric trace-free representation of the rotation group SO(3)[63] and $p_{2m}$ denote the Laplace expansion coefficients of the fiber orientation distribution function (fODF) normalized such that the fODF integrates to 1. This normalization differs from our previous study[9] by a factor of $1/5$, and is chosen for later convenience. Notice that only $p_{2m}$ and $l=2$ spherical harmonics $Y_2^m$, corresponding to the truncated Laplace expansion of the fODF, appear in Eq. (23). If $\Delta\bar{\chi}=0$, our previously published result for dispersed cylinders with scalar susceptibility[9] is obtained. The orientation dependence of the mesoscopic contribution to $\mathbf{L}^C$ proportional to $\Delta\bar{\chi}$ (second term in Eq. (21)) is embodied by a tensor with non-zero trace, in contrast to the tensor associated with the $\bar{\chi}^C$ contribution (first term in Eq. (21)), which has zero trace. When dispersion is uniform, the scatter matrix $\mathbf{T}$ is equal to the identity $\mathbf{T}=\mathbf{I}/3$, which nulls the contribution from $\bar{\chi}^C$ such that $\mathbf{L}^C = \mathbf{I}\Delta\bar{\chi}(1+\lambda)/18$. Hence, the model predicts a non-zero frequency shift even for uniform orientation dispersion (see Figure 4) due to $\Delta\bar{\chi}$.

Using Eqs. (8) and (19), the macroscopic contribution from the cylinders can be approximated on the scale of the MRI sampling resolution to be

$$\bar{\Omega}^{\text{Macro}}(\mathbf{R}) = \gamma B_0 \hat{\mathbf{B}}^{\text{T}} \sum_{\mathbf{R}'} \bar{\mathbf{Y}}(\mathbf{R}-\mathbf{R}')\left(\bar{\chi}^C(\mathbf{R}') + \frac{\Delta\bar{\chi}(\mathbf{R}')}{2}\left(\frac{1}{3}\mathbf{I}-\mathbf{T}(\mathbf{R}')\right)\right)\hat{\mathbf{B}}, \text{ (from cylinders). (24)}$$

Equation (24) describes how $\Delta\bar{\chi}$ gives rise to a macroscopic orientation dependence of the cylinders. The bulk susceptibility tensor, which STI aims to quantify[64], thus depend on the fODF of WM axons.

It is only in the limit of uniformly dispersed cylinders, the $\Delta \overline{\chi}$ term vanishes. This means that the apparent susceptibility anisotropy in the macroscopic contribution depends on the distribution of axons and the degree of orientation dispersion on the mesoscopic scale.

*Larmor frequency shift from spherical inclusions*

The contribution from spheres to the Larmor frequency is derived in supporting material S2. When the spheres occupy a low volume fraction, the Lorentzian tensor $\mathbf{L}^S$, Eq. (15), is found to be

$$\mathbf{L}^S = \frac{1}{2}\left(-\overline{\chi}^M + \left(\overline{\chi}^E + \overline{\chi}^A\right)\frac{\zeta^C}{\zeta^W}\right)\left(\mathbf{T} - \frac{1}{3}\mathbf{I}\right). \tag{25}$$

Equation (25) reflects the anisotropic mesoscopic contribution that emerges, because the reporting MR fluid and spherical inclusions are spatially restricted due to the presence of the cylinders. We only include contributions up to first order in sphere volume fraction in Eq. (25). However, $\mathbf{L}^S$ also includes contributions that is second order in sphere volume fraction. For high volume fraction of spheres, the full result is found in Eq. (S63) in supporting material. If we assume a volume fractions of axons around $\zeta^C \sim 0.35$ then the ratio $\zeta^C / \zeta^W \sim 0.5$. This means that positively magnetized extra-axonal (e.g. iron) and intra-axonal sources with susceptibility $\overline{\chi}^E + \overline{\chi}^A > 0$ can produce a mesoscopic frequency shifts on the same order of magnitude as $\overline{\chi}^C < 0$, if $-\overline{\chi}^C \sim \overline{\chi}^E + \overline{\chi}^A$. Hence, we cannot conclude that an orientation dependent Larmor frequency shift comes from the cylinders, as it just as well could reflect e.g. iron or intra-axonal sources[11,14]. The contribution from $\overline{\chi}^M$ (spheres in MW) in Eq. (25) generates a contribution to the Lorentzian tensor $\mathbf{L}^S$ similar to that of $\overline{\chi}^C$ to $\mathbf{L}^C$ in Eq. (21). Therefore, $\overline{\chi}^C$ can in practice be considered an effective susceptibility of both the cylinder layers (lipids) and spheres in MW (e.g. MBP).

The macroscopic contribution from spheres is

$$\overline{\Omega}^{\text{Macro}}(\mathbf{R}) = \gamma B_0 \sum_{\mathbf{R}'} \hat{\mathbf{B}}^T \overline{\mathbf{Y}}(\mathbf{R} - \mathbf{R}')\hat{\mathbf{B}}\, \overline{\chi}^S(\mathbf{R}'), \text{ (from spheres)}. \tag{26}$$

Here $\overline{\chi}^S$ defines the bulk susceptibility of all spherical inclusions, and does not depend on their anisotropic arrangement similar to $\overline{\chi}^C$ in Eq. (24).

*Total mesoscopic frequency shift $\bar{\Omega}^{\text{Meso}}$*

By combining Eqs. (21) and (25), the total mesoscopic contribution $\bar{\Omega}^{\text{Meso}}(\mathbf{R})$ to the mesoscopically averaged Larmor frequency shift $\bar{\Omega}(\mathbf{R})$, Eq. (2), from both cylinders and spheres becomes

$$\bar{\Omega}^{\text{Meso}}(\mathbf{R}) = -\left(\bar{\chi}^{\text{C}} + \bar{\chi}^{\text{M}} - \left(\bar{\chi}^{\text{E}} + \bar{\chi}^{\text{A}}\right)\frac{\varsigma^{\text{C}}}{\varsigma^{\text{W}}}\right)\frac{1}{2}\left(\hat{\mathbf{B}}^{\text{T}}\mathbf{T}\hat{\mathbf{B}} - \frac{1}{3}\right) \\ + \frac{\Delta\bar{\chi}}{12}\left(\hat{\mathbf{B}}^{\text{T}}\mathbf{T}\hat{\mathbf{B}}(1-\lambda) + \left(\lambda + \frac{1}{3}\right)\right) . \qquad (27)$$

Equation (27) describing the mesoscopic frequency shift from both cylinders and spheres is our main theoretical result in this study, and reduce to our previous results[9] when $\bar{\chi}^{\text{C}}$ is the only non-zero magnetic susceptibility.

*$\bar{\Omega}^{\text{Meso}}$ for axially symmetric fODF*

If the fODF is axially symmetric, only $Y_2^0(\hat{\mathbf{B}})$ contributes to the mesoscopic frequency shift in Eq. (27). In this case

$$\bar{\Omega}^{\text{Meso}} = a\sin^2(\theta_{\hat{\mathbf{B}}}) + b, \text{ (axially symmetric fODF)}, \qquad (28)$$

where

$$a = \frac{1}{2}\gamma\text{B}_0\left(\bar{\chi}^{\text{C}} + \bar{\chi}^{\text{M}} - \left(\bar{\chi}^{\text{E}} + \bar{\chi}^{\text{A}}\right)\frac{\varsigma^{\text{C}}}{\varsigma^{\text{W}}} - \frac{1}{6}\Delta\bar{\chi}(1-\lambda)\right)p_2 \qquad (29)$$

$$b = \gamma\text{B}_0\frac{1}{3}\left(-p_2\left(\bar{\chi}^{\text{C}} + \bar{\chi}^{\text{M}} - \left(\bar{\chi}^{\text{E}} + \bar{\chi}^{\text{A}}\right)\frac{\varsigma^{\text{C}}}{\varsigma^{\text{W}}}\right) + \Delta\bar{\chi}\frac{1}{6}\left((1+\lambda) + p_2(1-\lambda)\right)\right). \qquad (30)$$

The angle $\theta_{\hat{\mathbf{B}}}$ is the angle between the field and symmetry axis of the fODF. Here

$$p_2 = \sqrt{\frac{1}{20\pi}\sum_{m=-2}^{2}|p_{2m}|^2} \in [0,1] \qquad (31)$$

is the $l = 2$ rotation invariant of the fODF [65,66], which serves as a natural attenuation factor for the orientation dependent $\sin^2(\theta_{\hat{\mathbf{B}}})$ term due to orientation dispersion. This is qualitatively similar but quantitatively different from the fractional anisotropy (FA) of the diffusion tensor, which has been proposed in previous studies[23].

Figure 4A-B shows $\bar{\Omega}^{\text{Meso}} / \gamma B_0 |\bar{\chi}^C|$ only from cylinders with orientations following a Watson fODF[67]. Here the volume fraction is $\zeta^C = 0.3$, $\zeta^W = 0.7$, $g = 0.65$, $d/(d + d^W) = 2/3$ and $\zeta^{AW} = 0.2$ resulting in $\lambda = 1.6$. Figure 4A demonstrates the effect of increased dispersion when $\Delta \bar{\chi} = -\bar{\chi}^C / 5$, as chosen based on previous findings[23]. Dispersion is here quantified by a dispersion angle[65] $\theta_{p_2}$ ranging from 0 degrees up to the magic angle $\theta_{ma} \simeq 54.7$ degrees, where the dispersion effect will be maximum. Here, a non-zero frequency shift remains present due to non-zero susceptibility anisotropy $\Delta \bar{\chi}$ in the limit of fully dispersed cylinders ($p_2 = 0$). Figure 4B shows the effect when varying $\Delta \bar{\chi} / \bar{\chi}^C$, where the main effect is to shift of $\bar{\Omega}^{\text{Meso}} / \gamma B_0 |\bar{\chi}^C|$ up or down. The effect on the slope depends on how much $\lambda$ differs from 1, cf. Eq. (29). Figure 4C shows $\bar{\Omega}^{\text{Meso}} / \gamma B_0 |\bar{\chi}^C|$ for a nonzero spherical susceptibility $(\bar{\chi}^A + \bar{\chi}^E)/\bar{\chi}^C$. The slope of $\bar{\Omega}^{\text{Meso}} / \gamma B_0 |\bar{\chi}^C|$ changes substantially when the spherical susceptibility is comparable to $|\bar{\chi}^C|$.

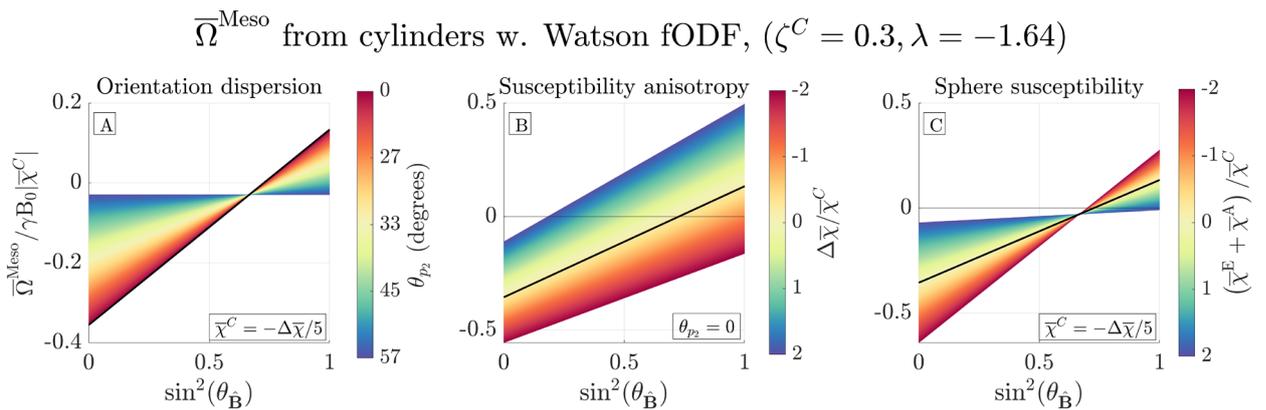

**Figure 4 – Mesoscopic frequency shift $\bar{\Omega}^{\text{Meso}}$ from cylinders and spheres with axially symmetric dispersion and anisotropic susceptibility: A** shows $\bar{\Omega}^{\text{Meso}}$ only for cylinders, with susceptiblity described by the legend and for varying angles $\sin^2(\theta_{\hat{\mathbf{B}}})$ between the external field and fODF symmetry axis. Colors represent the dispersion angles. As the

dispersion increases, the frequency variation are dampened with a fixed focal point at the magic angle. **B** shows variation in the ratio between the isotropic $\bar{\chi}^C$ and anisotropic magnetic susceptibility $\Delta\bar{\chi}$, indicated by the colors. Here, the dispersion is set to 0 (parallel cylinders). $\Delta\bar{\chi}$ mainly shifts the Larmor frequency up or down. **C** shows the change in the mesoscopic frequency for a non-zero spherical magnetic susceptibility $\bar{\chi}^E + \bar{\chi}^A$. Here the spherical susceptibility changes the slope but keeps the focal point fixed at the magic angle. The black lines in **A-C** indicate the same configuration across figures.

## $\bar{\Omega}^{\text{Meso}}$ for a non-axially symmetric fODF

To understand the effect of having a non-axially symmetric fODF, we assume for simplicity susceptibility anisotropy is zero, $\Delta\chi = 0$, in this section. Consider first Eq. (28) for an axially symmetric fODF, where $\Delta\chi = 0$ implies $a = \gamma B_0 \bar{\chi}_{\text{eff}} p_2 / 2$ and $b = -\gamma B_0 \bar{\chi}_{\text{eff}} p_2 / 3$. Here, $\bar{\chi}_{\text{eff}}$ denotes the effective susceptibility $\bar{\chi}_{\text{eff}} = \left( \bar{\chi}^C + \bar{\chi}^M - \left( \bar{\chi}^E + \bar{\chi}^A \right) \zeta^C / \zeta^W \right)$. Hence, $\bar{\Omega}^{\text{Meso}}\left(\theta_{\hat{\mathbf{B}}}\right) = \gamma B_0 \bar{\chi}_{\text{eff}} p_2 (\sin^2(\theta_B)/2 - 1/3)$ will always be zero when $\theta_{\hat{\mathbf{B}}}$ is equal to the magic angle $\theta_{\text{MA}} \simeq 54.7°$. This is demonstrated in Figure 5A for the Watson distribution[67] with $\hat{\mathbf{z}}$ as symmetry axis and with dispersion angle [68] $\theta_{p_2}$ ranging from 5.7 degrees to $\theta_{\text{MA}}$. All curves cross 0 (black dotted line) at $\theta_{\text{MA}}$, while increasing dispersion attenuates the slope of the frequency shift vs. $\sin^2\theta_B$. To examine the effect of having a non-axially symmetric fODF, the Bingham distribution[67] with dispersion axes along $\hat{\mathbf{x}}$ and $\hat{\mathbf{y}}$ was used as fODF in Figure 5B-D. The dispersion angle $\theta_{\hat{x}}$ along $\hat{\mathbf{x}}$ is fixed at 5.7 degrees, while dispersion $\theta_{\hat{y}}$ is varied from 5.7 degrees to $\theta_{\text{MA}}$. When the dispersions along $\hat{\mathbf{x}}$ and $\hat{\mathbf{y}}$ are equal, the Bingham fODF reduces to the Watson fODF. Since the trace of the scatter matrix is unity, $\text{Tr}[\mathbf{T}] = 1$, increasing dispersion along $\hat{\mathbf{y}}$ by an angle $\theta_{\hat{y}}$ results in increasing the eigenvalue of $\mathbf{T}$ corresponding to the eigenvector along $\hat{\mathbf{y}}$, while the corresponding eigenvalue to $\hat{\mathbf{z}}$ decreases due to conservation of trace of $\mathbf{T}$ (since dispersion along $\hat{\mathbf{x}}$ is small, we can neglect any change in its eigenvalue of $\mathbf{T}$ after varying $\hat{\mathbf{y}}$). In Figure 5B, the field is rotated in the $\hat{\mathbf{x}}\hat{\mathbf{z}}$ plane. Here the focal point appears at 90 degrees, and the zero-frequency angle now becomes a decreasing function of dispersion. In Figure 5C, the field is at an equal azimuthal angle to the two dispersion axes which makes the frequency shift appear axially symmetric, as the zero-crossing is at $\theta_{\text{MA}}$ again.

Finally, in Figure 5D, the field is tilted in the $\hat{\mathbf{y}}\hat{\mathbf{z}}$ plane. The zero-frequency angle is now shifted to the right as both the frequency slope and offset are reduced. As a consequence, this means that non-axially symmetric dispersion can mimic the effect of susceptibility anisotropy, and only by knowing the fODF it is possible to discern the actual susceptibility anisotropy.

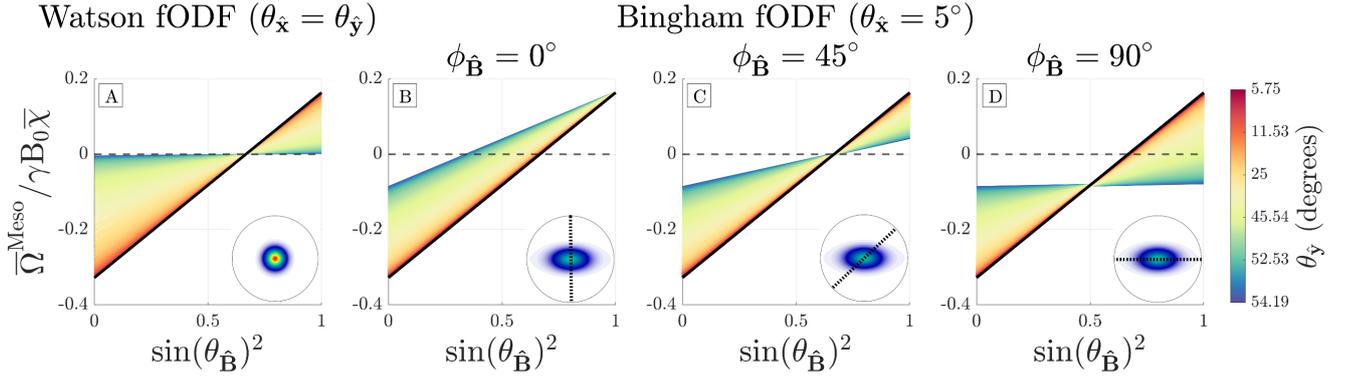

**Figure 5 – Mesoscopic frequency shift from cylinders with orientation disperson and scalar susceptiblity: A** describes $\overline{\Omega}^{\text{Meso}}$ for an axially symmetric Watson distribution, as a function of the angle $\theta_{\hat{\mathbf{B}}}$ between the applied field and the symmetry axis of the cylinders' fODF. The colors encode the dispersion angle, as indicated by the colorbar to the right. As the dispersion increases, the mesoscopic frequency is dampened. **B-D** shows $\overline{\Omega}^{\text{Meso}}$ for a non-axially symmetric Bingham distribution with axes $\hat{\mathbf{x}}$ and $\hat{\mathbf{y}}$, and with a fixed dispersion angle $\theta_{\hat{\mathbf{x}}} = 5°$, while the colors encode the dispersion angle $\theta_{\hat{\mathbf{y}}}$. B shows the mesoscopic frequency vs. tilt angle $\theta_{\hat{\mathbf{B}}}$ in the zx-plane ($\phi_{\hat{\mathbf{B}}} = 0°$) indicated by the dashed line (--), **D** in the zy-plane ($\phi_{\hat{\mathbf{B}}} = 90°$), while **C** is in between ($\phi_{\hat{\mathbf{B}}} = 45°$). The black lines indicate the same fODF across all four figures.

## 3 | Methods

**Simulations**

We designed two sets of simulations to **(a)** validate the analytical result, Eq. (21), and to **(b)** demonstrate the feasibility of solving the inverse problem and extract susceptibility parameters from frequency measurements. All simulations were done in MATLAB (the MathWorks, Natick, MA, USA). The data that support the findings of this study are available from the corresponding author upon reasonable request.

*Simulation (a): Validity of the Lorentzian tensor for a distribution of cylinders with susceptibility anisotropy $\Delta\chi$*

A total of 11 unique samples of non-overlapping cylindrical shells with varying orientation dispersion were constructed to validate the model. The sample is represented by a cubic volume with side lengths $L$. Figure 6 gives an overview of the sample geometries. The radii of the cylinders $\rho$ were drawn from a gamma distribution with mean and standard deviation (SD) $0.0143L$, and the g-ratio $g$, the ratio between the inner and outer radii, was fixed at $g = 0.65$ [69]. The positions were drawn from a uniform distribution. The total volume fraction achieved this way was $\zeta \approx 0.15$, and $\lambda \approx 2.2$. Dispersion was generated from an axially symmetric and uniform distribution of orientations $\hat{\boldsymbol{n}}$, within an allowed polar-range $\cos^{-1}(\hat{\boldsymbol{z}} \cdot \hat{\boldsymbol{n}}) = \theta < \theta_c$, where $\theta_c$ is the cut-off angle. Every sample was discretized in a 3D grid, with a resolution set by the number of grid points. This defined the indicator function $v^C(\boldsymbol{r})$ of the cylinders on the discretized grid. The susceptibility tensor $\boldsymbol{\chi}^C(\boldsymbol{r})$ was generated using Eq. S6 in supporting material. This required calculating the azimuthal angles in local coordinates and orientation for each cylinder on the 3D grid to compute the susceptibility tensor in the lab frame. Susceptibilities were chosen to be maximally anisotropic by setting $\chi^C = \frac{1}{3}\Delta\chi$, c.f. Eq. (8), as previous simulations[9] already validated the isotropic contribution. Using a discrete Fourier-transform with periodic boundaries, $v^C(\boldsymbol{k})$, $\boldsymbol{\chi}^C(\boldsymbol{k})$, and in turn the correlation tensor $\boldsymbol{\Gamma}(\boldsymbol{k})$, Eq. S2 in supporting material, were computed. From this, the normalized Lorentzian tensor $\mathbf{L}/(\zeta\Delta\chi(\lambda+1)/6)$ could be estimated solely from susceptibility anisotropy using Eq. S1, and compared to the model prediction $\frac{1}{2}(\mathbf{I} - \mathbf{T})$, cf. Eq. (17). The eigenvalues $(\lambda_{\perp 1}, \lambda_{\perp 2}, \lambda_{\parallel})$ of $\mathbf{L}/(\zeta\Delta\chi(1+\lambda)/6)$ were calculated for both simulation and model prediction and compared for each population.

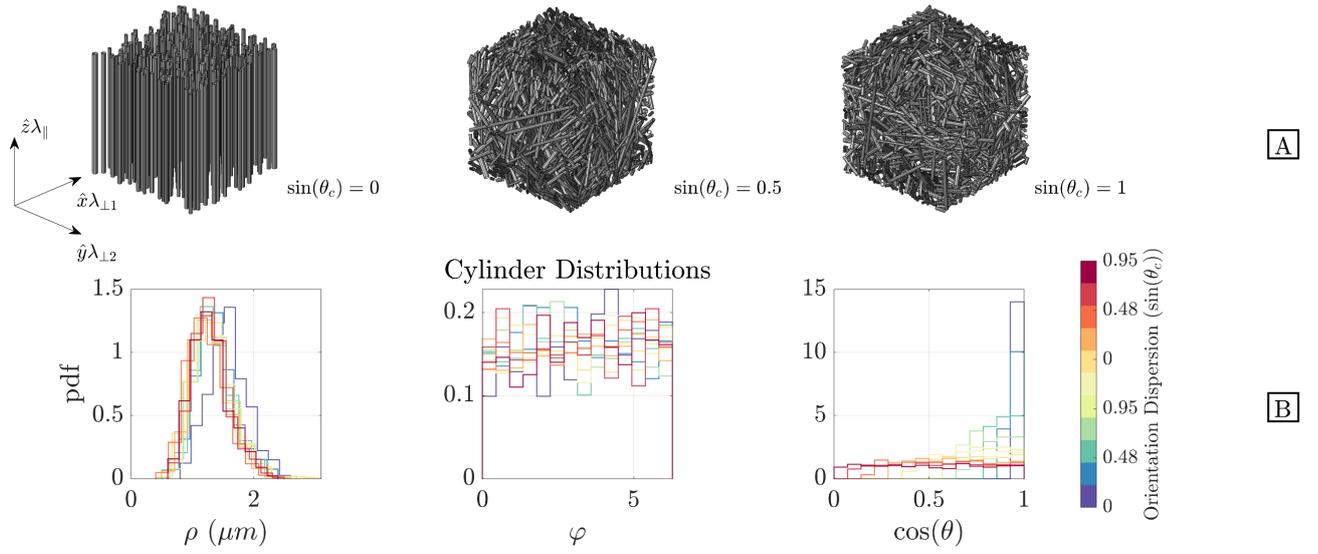

**Figure 6 – simulation (a):** Populations of cylinders with different levels of orientation dispersion are shown in **A**. **B** shows the probability density function (pdf) of the resulting cylinder parameters for each configuration. The cylinder radius $\rho$ is gamma-distributed, while $\theta$ and $\varphi$ are uniformly distribution in the full range of azimuthal angle and from zero to the maximum polar angle $\theta_c$, respectively. Colors are used to represent different populations with orientation dispersion indicated by the color bar.

*Simulation (b): Feasibility of solving inverse problem*

Estimating the magnetic susceptibility $\vec{\bar{\chi}}$ from the MRI measured Larmor frequency shift amounts to solving the linear system of equations $\vec{\Omega}_{MRI} = \mathbf{A}\vec{\bar{\chi}}$ (here vectors indicate a vectorization over the voxels):

$$\begin{pmatrix} \vec{\Omega}_{MRI}(\hat{\mathbf{B}}_1) \\ \vec{\Omega}_{MRI}(\hat{\mathbf{B}}_2) \\ \vdots \\ \vec{\Omega}_{MRI}(\hat{\mathbf{B}}_n) \end{pmatrix} = \gamma B_0 \begin{pmatrix} \mathbf{A}^{\chi^C}(\hat{\mathbf{B}}_1) & \mathbf{A}^{\Delta\chi}(\hat{\mathbf{B}}_1) & \mathbf{A}^{\lambda\Delta\chi}(\hat{\mathbf{B}}_1) & \mathbf{A}^{\chi^S}(\hat{\mathbf{B}}_1) \\ \mathbf{A}^{\chi^C}(\hat{\mathbf{B}}_2) & \mathbf{A}^{\Delta\chi}(\hat{\mathbf{B}}_2) & \mathbf{A}^{\lambda\Delta\chi}(\hat{\mathbf{B}}_2) & \mathbf{A}^{\chi^S}(\hat{\mathbf{B}}_2) \\ \vdots & \vdots & \vdots & \vdots \\ \mathbf{A}^{\chi^C}(\hat{\mathbf{B}}_n) & \mathbf{A}^{\Delta\chi}(\hat{\mathbf{B}}_n) & \mathbf{A}^{\lambda\Delta\chi}(\hat{\mathbf{B}}_n) & \mathbf{A}^{\chi^S}(\hat{\mathbf{B}}_n) \end{pmatrix} \begin{pmatrix} \vec{\bar{\chi}}^C \\ \Delta\vec{\bar{\chi}} \\ \lambda\Delta\vec{\bar{\chi}} \\ \vec{\bar{\chi}}^S \end{pmatrix} \quad (32)$$

The vector $\vec{\bar{\chi}} = \begin{pmatrix} \vec{\bar{\chi}}^C & \Delta\vec{\bar{\chi}} & \lambda\Delta\vec{\bar{\chi}} & \vec{\bar{\chi}}^S \end{pmatrix}^T$ is a $4N \times 1$ block vector in voxel space containing the 4 main bulk susceptibilities, where N is the number of voxels. $\vec{\Omega}_{MRI}$ has dimensions $nN \times 1$ for $n$ unique sampling orientations $\hat{\mathbf{B}}$. Each $\mathbf{A}^{(i)}(\hat{\mathbf{B}})$ is an $N \times N$ matrix in voxel space where every element includes a tensor contraction with $\hat{\mathbf{B}}$ describing the magnetic response, which is here referred to as magnetic space to distinguish the two. $\bar{\chi}^S$ is assumed to be small such that its mesoscopic contribution, Eq. (25), can be neglected. The elements $A^{(i)}(\mathbf{R}_i, \mathbf{R}_j)$ of every block matrix $\mathbf{A}^{(i)}(\hat{\mathbf{B}})$, for a row in the response matrix $\mathbf{A}$, can be represented as

$$A^{\chi^C}(\mathbf{R}_i, \mathbf{R}_j) = M(\mathbf{R}_i)\hat{\mathbf{B}}^T \left[ -\frac{1}{2}T(\mathbf{R}_j) + \frac{1}{3}I + \bar{Y}(\mathbf{R}_i - \mathbf{R}_j) \right] \hat{\mathbf{B}} M(\mathbf{R}_j)$$

$$A^{\lambda\Delta\chi}(\mathbf{R}_i, \mathbf{R}_j) = \frac{1}{12} M(\mathbf{R}_i)\hat{\mathbf{B}}^T \left[ I - T(\mathbf{R}_j) \right] \hat{\mathbf{B}} M(\mathbf{R}_j)$$

$$A^{\Delta\chi}(\mathbf{R}_i, \mathbf{R}_j) = -\frac{1}{2} M(\mathbf{R}_i)\hat{\mathbf{B}}^T \left[ \bar{Y}(\mathbf{R}_i - \mathbf{R}_j)T(\mathbf{R}_j) - \frac{1}{3}\bar{Y}(\mathbf{R}_i - \mathbf{R}_j) - \frac{1}{6}T(\mathbf{R}_j) - \frac{1}{18}I \right] \hat{\mathbf{B}} M(\mathbf{R}_j)$$

$$A^{\chi^S}(\mathbf{R}_i, \mathbf{R}_j) = M(\mathbf{R}_i)\hat{\mathbf{B}}^T \bar{Y}(\mathbf{R}_i - \mathbf{R}_j) \hat{\mathbf{B}} M(\mathbf{R}_j)$$

(33)

Here $M(\mathbf{R}_i)$ a diagonal mask matrix describing the spatial distribution of allowed susceptibility sources and where $\vec{\Omega}_{\text{MRI}}$ is successfully measured[70] (here the sample mask). The matrix dimension of the total response matrix $\mathbf{A}$ is far too great to physically allocate, which hampers the possibility to solve Eq. (32) in closed form. However, $\mathbf{A}$ implements fast linear operations such as convolutions and simple matrix multiplications. For these types of linear problems of impractically high dimensionality, one may turn to iterative least squares schemes, which has the benefit of never allocating any matrices, but declares every matrix as a function handle, so only vectors of dimension equal to $\vec{\Omega}_{\text{MRI}}$ and $\vec{\chi}$ are allocated. An example of such solvers is the LSMR [71] algorithm, which is used here to invert Eq. (32). Using these simulations, we tested the feasibility of extracting susceptibility parameters from whole brain measurements of the MRI frequency shift $\bar{\Omega}_{\text{MRI}}(\hat{\mathbf{B}})$ acquired at multiple sample orientations $\hat{\mathbf{B}}$. This was done by generating a digital brain phantom from previously acquired data[24]. The phantom is outlined in Figure 7 and based on dMRI measurements of an ex vivo mouse brain at 16.4T. The brain was measured with a matrix size of 102x102x170 and was segmented into gray and white matter from $b = 0$ ms/µm² images using SPM[72]. From DKI[73] fitting (b=3, 5 ms/µm², 30 dir.), FA and mean diffusivity (MD) was extracted from the diffusion tensor. Laplace expansion coefficients of the fODF, $p_{2m}$, were estimated using FBI[74] (b=10ms/µm², 75 dir.). From these, maps of 4 orientationally invariant parameters were synthesized: $\bar{\chi}^C$, $\Delta\bar{\chi}$, $\lambda\Delta\bar{\chi}$ and $\bar{\chi}^S$. All 4 parameters would subsequently need to be estimated in each voxel of the whole brain. Here $\bar{\chi}^S = \bar{\chi}^S_{WM} + \bar{\chi}^S_{GM}$ are the bulk susceptibilities from spheres in gray matter (GM) and WM. The resulting Larmor frequencies from each of these "sources" are $\bar{\Omega}^{Meso}_{\bar{\chi}^C}(\hat{\mathbf{B}}) + \bar{\Omega}^{Macro}_{\bar{\chi}^C}(\hat{\mathbf{B}})$, $\bar{\Omega}^{Meso}_{\Delta\bar{\chi}}(\hat{\mathbf{B}}) + \bar{\Omega}^{Macro}_{\Delta\bar{\chi}}(\hat{\mathbf{B}})$, $\bar{\Omega}^{Meso}_{\Delta\bar{\chi}\lambda}(\hat{\mathbf{B}})$ and $\bar{\Omega}^{Macro}_{\bar{\chi}^S}(\hat{\mathbf{B}})$, and computed according to Eq. (32). Their sum defines the total MRI Larmor frequency shift $\bar{\Omega}_{MRI}(\hat{\mathbf{B}})$ for each orientation $\hat{\mathbf{B}}$. Figure 7 shows the susceptibility and noiseless frequency maps. The inverse problem was solved for different numbers of sample orientations made using an electrostatic repulsion scheme[75], adding to $\bar{\Omega}_{MRI}(\hat{\mathbf{B}})$ different levels of Gaussian noise $\varepsilon \sim N(0, \sigma^2 / \text{SNR}^2)$ by varying the signal-to-noise ratio (SNR), and lastly, the maximum polar angle of sample rotation. The solution with the lowest root-mean-squared error (RMSE) compared to ground truth during fitting was chosen for analysis.

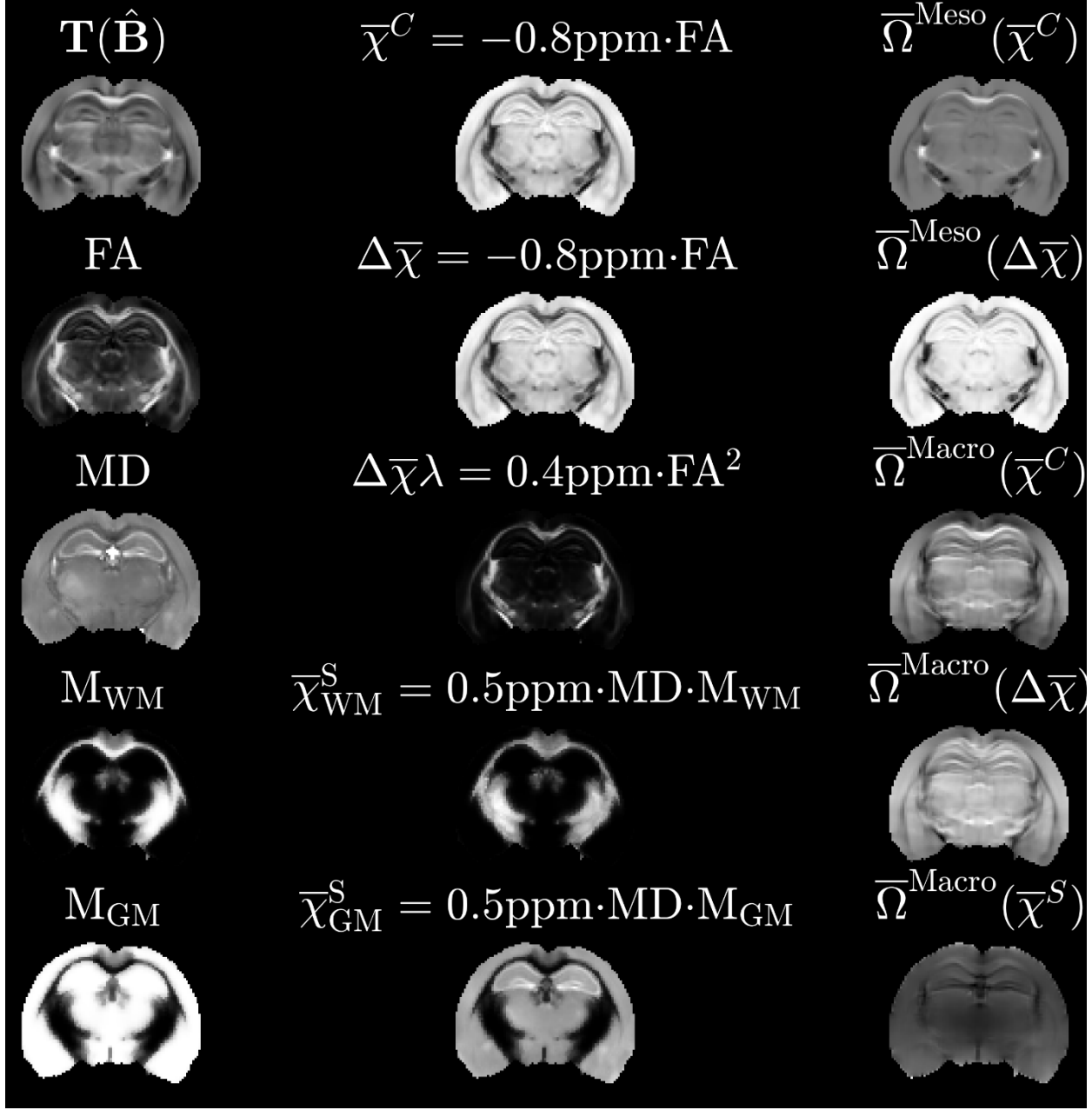

**Figure 7 – Simulation (b): Digital susceptibility phantom.** FA, MD, and fODF from dMRI measurement of an ex vivo mouse brain (left column). White and gray matter masks, $M_{WM}$ and $M_{GM}$ respectively, were extracted using SPM [72]. From this we defined ground truth susceptibility maps (middle column). Right column shows corresponding noiseless Larmor frequencies for a particular sample orientation $\hat{\mathbf{B}}$. The sum of all frequencies defines the MRI measured

frequency shift $\bar{\Omega}_{MRI}$. The aim is to estimate the susceptibility maps from the measured frequency $\bar{\Omega}_{MRI}$ acquired at multiple sample orientations and after adding noise.

# 4 | Results

**Simulations:**

*Simulation (a): Validity of the Lorentzian tensor for a distribution of cylinders with susceptibility anisotropy*

Figure 8 shows the eigenvalues $(\lambda_{\perp 1}, \lambda_{\perp 2}, \lambda_{\parallel})$ of $\mathbf{L}/(\zeta\Delta\chi(1+\lambda)/6)$ and $\frac{1}{2}(\mathbf{I}-\mathbf{T})$ from all the cylinder samples with varying orientation dispersion, where $\chi^C = \Delta\chi/3$. The model predicts the same behavior as the ground truth simulation. In contrast to our previous results[9] for the Lorentzian tensor relating $\chi^C$, here the Lorentzian tensor from susceptibility anisotropy $\Delta\chi$ has non-zero eigenvalues for all levels of orientation dispersion. The eigenvector for $\lambda_{\parallel}$, corresponding to the symmetry axis of the cylinder orientation distribution, also agreed with simulations within a mean angular error of 3.5 degrees (data not shown).

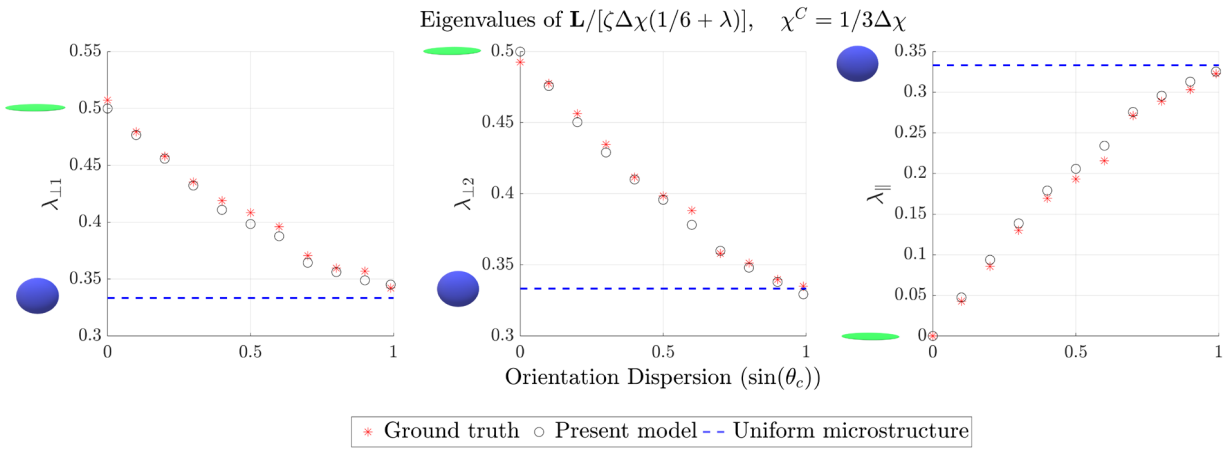

**Figure 8 - Simulation (a):** Simulation of the mesoscopic contribution from 11 different orientation distributions. Eigenvalues $(\lambda_{\perp 1}, \lambda_{\perp 1}, \lambda_{\parallel})$ of ground truth $\mathbf{L}/(\zeta\Delta\chi(1+\lambda)/6)$ and model prediction $(\mathbf{I}-\mathbf{T})/2$ are presented for various levels of dispersion set by the maximum allowed polar angle $\theta_c$, where $\chi^C = \Delta\chi/3$. Notice the trace of $\mathbf{L}/\left(\zeta\Delta\chi(1+\lambda)/6\right)$ is 1. When $\theta_c = 0$, $\bar{\Omega}^{\text{Meso}}$ is non-zero only perpendicular to the cylinder, described by the green pancake. When $\theta_c = 1$, $\bar{\Omega}^{\text{Meso}}$ is non-zero and isotropic, described by the blue sphere. Uniform microstructure defines the limit of uniform orientation dispersion.

*Simulation (b): Feasibility of solving inverse problem*

Figure 9 shows the resulting fits and normalized RMSE for each of the four susceptibilities (each normalized by the difference between maximum and minimum of the respective ground truth susceptibilities). At an SNR=50-100, all parameters were below 12% RMSE across all numbers of orientations. Computation time was around 10 minutes on an off-the-shelf laptop. Reasonable fitting accuracy was achieved for 7 orientations and realistic frequency SNR. Furthermore, while decreasing the maximum tilt angle increased the RMSE, it did not completely erode the accuracy (still within 12% for SNR=100 and maximum angle of 45 degrees). Figure S7 and Figure S8 in supplementary material shows the resulting fits for varying tilt angle and sample orientations.

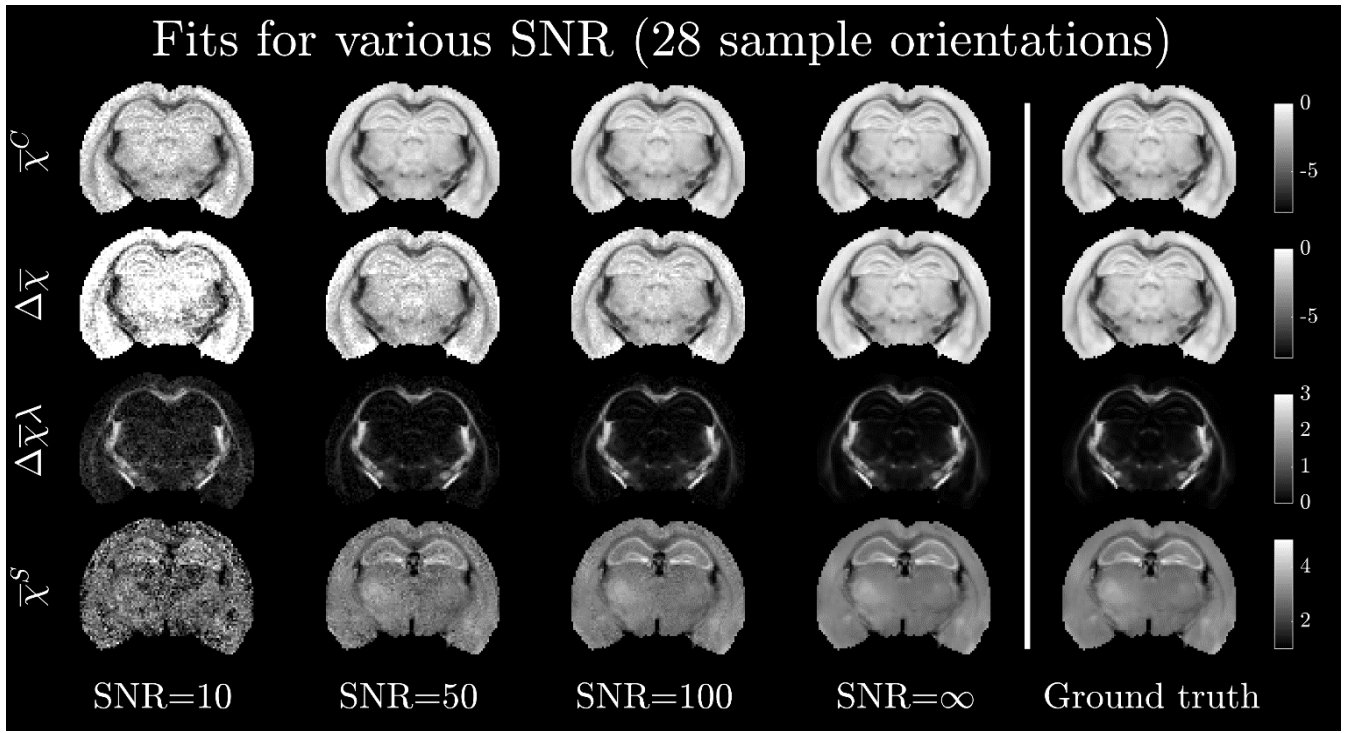
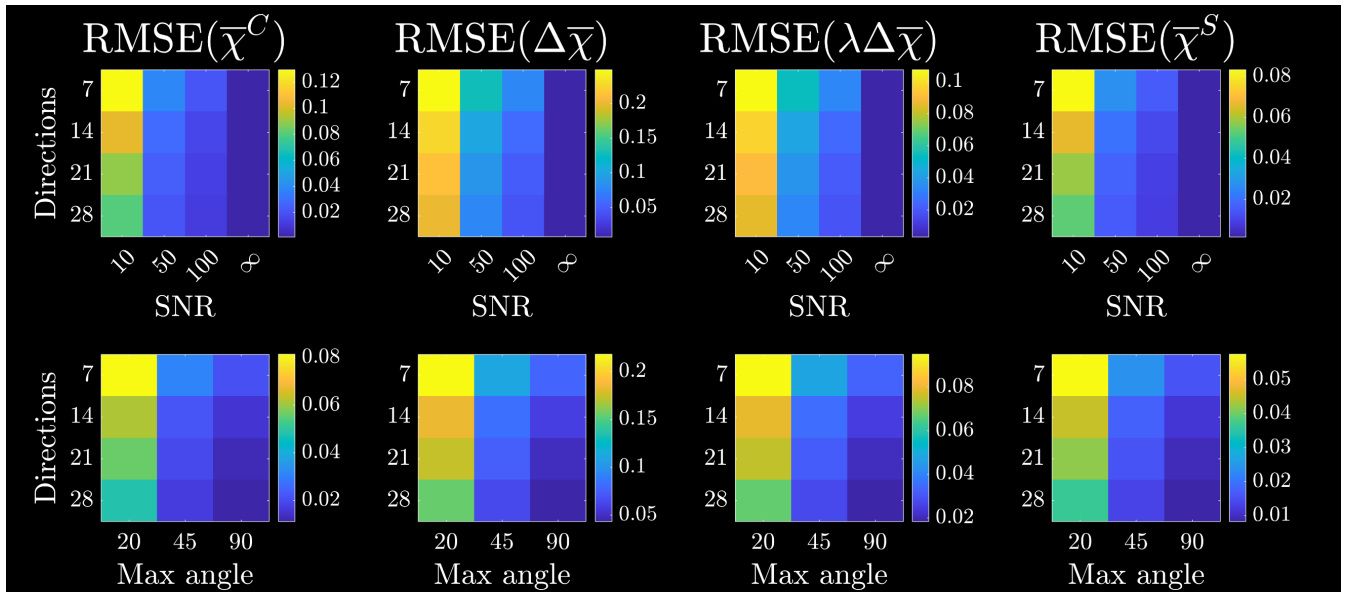

**Figure 9 – Simulation (b):** Susceptibility fitting maps. Fitting maps for various phase SNR, and 28 directions. RMSE: RMSE for each parameter normalized by the range of susceptibility in each ground truth parameter. The first row shows variation in SNR, while the latter shows variation in tilt angle for a fixed SNR=100.

# 5 | Discussion

**Implications of biophysical model of white matter on susceptibility estimation**

Identifying the relevant biophysical parameters needed to characterize susceptibility-induced Larmor frequency shifts in MRI has been an active field of research for many years, and has led to many insightful biophysical models[11,13,14,30] of white matter (WM) magnetic microstructure. However, despite this development, an ever-growing majority of QSM/STI experiments still effectively relies on the simplified description of tissue as a homogenous liquid[76]. While such model framework provides a simple analytical description and plausible looking images, it has the unfortunate downside of removing a potentially large specificity to magnetic microstructure, which inevitably biases susceptibility estimation[23], and could prevent a deeper understanding of tissue's magnetic composition[77].

Motivated by previous susceptibility models[9,11,13,14,24,30,32] and diffusion models[5] which have included different features of microscopic structural anisotropy and microscopic susceptibility anisotropy, this study aimed to take another step towards a realistic description of magnetic microstructure. We modeled WM axons as a dense media of long multi-layered cylinders with arbitrary orientation dispersion and microscopic susceptibility anisotropy originating from lipid chains forming the myelin sheaths. It further included spherical inclusions to model tissue iron and other similar contributors.

The mesoscopic contribution, Eq. (27), depends in general on the scatter matrix $\mathbf{T}$ which can be written as a linear combination of $l=2$ spherical harmonics $Y_2^m$. However, the functional behavior described by $a\sin^2(\theta_{\hat{\mathbf{B}}}) + b$ in Eq. (28), where $\theta_{\hat{\mathbf{B}}}$ is the angle between the axons and magnetic field, has been observed in many experiments[8,23,26,27]. One may then be led to think that the fODF is axially symmetric, but typically these experiments only rotate the sample around a single fixed axis, which can give the impression of an axially symmetric fODF. Second, the fact that the frequency shift is non-zero at the magic-angle has often been used as evidence of susceptibility anisotropy $\Delta\chi$, in accordance with the behavior for an axially symmetric fODF as seen in Figure 4A-B. However, as shown in Figure 5, a shift in the zero crossing can also be induced even for $\Delta\chi = 0$ when the fODF is not axially symmetric and the sample only rotated in a plane. This degeneracy becomes apparent when performing 3D rotations and can only be resolved by estimating orientation dispersion separately, e.g. with dMRI, which in turn will reduce a potential bias in parameter estimation of susceptibility anisotropy when the fODF is not axially symmetric. As the fODF also affects the macroscopic contribution from neighboring voxels, the only contribution accounted for in STI[22], we recommend estimating the fODF to obtain proper measurements of susceptibility anisotropy of WM axons.

Equation (25), describing the Lorentzian tensor $\mathbf{L}^S$ caused by the spherical magnetic susceptibility $\bar{\chi}^S$, shows the importance of considering a dense media of magnetic microstructure. Here a substantial mesoscopic contribution $\bar{\Omega}^{\text{Meso}}$ from spheres is caused by a high density of cylinders reducing the available space for water molecules in an anisotropic way. This result may be particularly important when studying superficial WM[38,39], which contains large amount of iron in the extra-axonal space.

Previous models[9,11,24,32] describing the mesoscopic frequency shift $\bar{\Omega}$ from WM axons relied on the ability to write the Lorentzian tensor as $\mathbf{L} = -\mathbf{N}\chi$, i.e. as a product between a susceptibility tensor $\chi$ and a purely structural tensor, here denoted as $\mathbf{N}$. However, as shown by Eq. (3), $\mathbf{L}$ depends in general on the tensor-valued magneto-structural correlation function $\Gamma(k)$. A relevant question is thus if and when $\mathbf{L}$ can be simplified to a product $\mathbf{N}\chi$. Looking at the results in Eq. (21), it is clear that the intra-axonal water, here appearing through $\lambda$, adds an additional frequency shift that cannot be derived from assuming a constant (scalar or tensor) magnetic susceptibility of the cylinders. This can be concluded from previous results showing that when the susceptibility of the cylinder is constant, only structural correlations matter, and the mesoscopic frequency shifts in the intra- and extra-cylindrical compartments are the same[24]. If, however, the intra-cylindrical water is set to zero, such that the cylinders can be regarded as being solid, $\lambda = 0$ in Eq. (21). In this case, the mesoscopic frequency shift generated by every single solid cylinder can in fact be described by a product on the form $\mathbf{N}(\hat{n})\bar{\chi}(\hat{n})$, where $\bar{\chi}(\hat{n}) = \bar{\chi}^C \mathbf{I} - \Delta\bar{\chi}(\hat{n}\hat{n}^T - \mathbf{I}/3)/2$ is the mean magnetic susceptibility (cf. Eq. (19)) and $\mathbf{N}(\hat{n}) = (\hat{n}\hat{n}^T - \mathbf{I}/3)/2$ the demagnetization tensor[9] for a cylinder pointing along $\hat{n}$. However, upon summing over multiple cylinders with different orientations, it is in general not possible to write $\mathbf{L}$ as a product. The only exception to this statement is when axons are oriented axially symmetric around some $\hat{n}$, in which case $\mathbf{L}$ is a product of the form $p_2 \mathbf{N}(\hat{n})\bar{\chi}(\hat{n})$.

*Limitations and Future model extensions*

The proposed WM model is valid when the signal is adequately described by the first signal cumulant $\overline{\Omega}_{MRI} t$, which is the case in the diffusion narrowing regime (DNR) (see the discussion in our previous study[9]). While the first cumulant is also sufficient in the weak static dephasing regime, the measured signal $S(t)$ would likely depend on myelin water (MW) frequency shifts and other fast relaxing water pools, which are still not fully understood[53,78]. Myelin water may also affect the frequency shift through exchange with the surrounding water. The approximation of the signal in terms of the first cumulant can be violated for different reasons. it is violated in regions with high levels of iron or calcium, where the signal averaging is not equivalent to phase averaging[36]. Clustering of iron in neuroglia or within the axonal water compartments may also be of importance, but is so far less understood compared to GM[79,80]. We therefore plan to investigate WM iron in the future and extend our model in the future to account for any additional frequency effects. Second, it is well known that water is compartmentalized in WM, which can be seen from dMRI experiments[81,82]. Hence in DNR, the tissue is expected to have many intra-axonal compartments and extra-axonal compartments - even in fixated tissue. Luckily, when the phase is small, which is a main assumption in this study, one can perform a cumulant expansion of the compartmentalized signal, which yields: $S(t) = \exp\left(-i\overline{\Omega}_{MRI} t - R_2 t + \mathcal{O}(\delta\Omega^2 t^2)\right)$. Here, the exponent becomes a power series in $\delta\Omega t$ corresponding to the typical phase shift and the linear term is the first cumulant $\overline{\Omega}_{MRI}$ of the whole signal (cf. Eq. (1)). Nevertheless, analyzing the individual compartment's signals could be used to gain a deeper insight into the mesoscopic frequency shifts from different susceptibility sources, including their relaxation rates and water fractions[13,57,60,83,84], which can change the sensitivity to different susceptibility sources in different compartments, e.g., extra-axonal iron will affect intra-axonal water differently than extra-axonal water etc. This will be pursued in depth in future studies. Diffusion filtering can be used to change the relative weights of water compartments in the signal. For such techniques, our model framework can be used to describe frequency shifts from specific water compartments, e.g. intra-axonal water[85]. Incorporating the susceptibility model into diffusion models will be considered in future studies[86–90]. Recent susceptibility models[91] combines fitting of relaxation rates and frequency shifts to disentangle para- and diamagnetic sources in brain tissue. However, such models have so far disregarded mesoscopic frequency contributions, similarly to QSM. Nevertheless, the contrast from the estimated sources has shown great promise when compared to histology. It would therefore be of interest to expand the analytical framework presented here to derive an analytical solution for transverse relaxation.

While the analytical results have been validated numerically, the model remains to be validated against real WM tissue where axons exhibit beading, undulations, non-circular cross-sections, etc.[51,62,92]. Scrutinizing the model with realistic in-silico substrates acquired with 3D electron microscopy [51] will therefore be considered in future studies.

**Feasibility of solving inverse problem**

Simulation b) investigated the solution to the inverse problem in a digital brain phantom. The positive results suggest that successful model-estimation in humans in vivo, where maximum tilt angle and number of orientations are limiting factors, is feasible. The reason why susceptibility parameters can be estimated despite small tilt angles is that the tensor structure is already determined from diffusion. This leaves only 4 orientationally invariant susceptibility parameters to be determined in each voxel, if the mesoscopic frequency shift from spheres (cf. Eq. (25)) can be neglected. While this assumption could be violated in real WM, additional information about the axonal volume fraction may be needed to successfully solve the inverse problem. Alternatively, the mesoscopic frequency shift could be included in the inverse problem as $\overline{\Omega}^{\mathrm{Meso}}(\hat{\mathbf{B}}) = a\hat{\mathbf{B}}^{\mathrm{T}}\mathbf{T}\hat{\mathbf{B}} + b$, analogous to Eq. (28), but at the expense of being unable to separate $\overline{\chi}^C$ and $\overline{\chi}^S$. Qualitatively, the maps in Figure 9 still appear slightly noisy, especially $\overline{\chi}^S$ in GM. This makes sense as $\overline{\chi}^S$ in the phantom only has a macroscopic contribution $\overline{\Omega}^{\mathrm{Macro}}_{\chi^S}(\hat{\mathbf{B}})$ to the Larmor frequency $\overline{\Omega}_{\mathrm{MRI}}(\hat{\mathbf{B}})$, and is difficult to disentangle from $\overline{\chi}^C$ when $\overline{\Omega}^{\mathrm{Meso}}_{\chi^C}(\hat{\mathbf{B}})$ is close to zero (the two sources give rise to similar macroscopic frequency shifts). Weighted least squares or regularization may further help to decrease such degeneracy and reduce RMSE. Other limiting factors that may affect the fitting performance is noise in the fODF estimation, background field removal and unwrapping of the frequency shift[93], and finding an optimal stopping criteria (e.g., L-curve optimization [94–96]) for the fitting routine, when no ground truth is available. All such potential confounding factors will be explored in future studies. In simulation b), mesoscopic contributions from the spheres were disregarded. If this contribution is too large to ignore, it can make the inverse problem harder to solve. This is because the mesoscopic contribution from spheres is scaled by the cylinder volume fraction $\zeta^C$ in contrast to the macroscopic contribution, Eq. (22), which adds an additional degree of freedom to the inverse problem.

**Wharton and Bowtell residual frequency shift $f_R$**

Because of the comprehensive theoretical framework involved, we dedicated this work to a detailed presentation and analysis of the WM biophysical model for the Larmor frequency shift, leaving experimental validation as a future endeavor. However, an insightful study has already been performed by Wharton and Bowtell [23] in the spirit of previous experiments by Luo et al. [8] and Lee et al. [29]. In their study, a fresh porcine optical nerve was cast in agarose gel in a spherical container and scanned at multiple sample orientations to the scanner field. By measuring the frequency external to the optical nerve, it was possible to determine an average macroscopic frequency shift $\bar{\Omega}^{Macro}$ in a cylindrical shell surrounding the nerve. This was achieved with computer simulations by digitizing a sample mask in place of the sample, and then finding the optimal WM bulk susceptibility values $\left(\bar{\chi}^C, \Delta\bar{\chi}\right) = \left(\chi_I, \frac{3}{2}\chi_A\right)$ to describe the average frequency outside. In this way, the authors found $\chi_I = -0.082\,\text{ppm}$ and $\chi_A = 0.011\,\text{ppm}$. Next, they estimated a residual frequency shift $f_R = a\sin^2\left(\theta_{\hat{\mathbf{B}}}\right) + b$ with $a = -5.59\,\text{Hz}$ and $b = 4.88\,\text{Hz}$, which defined the remaining frequency shift inside the sample in addition to the one that could be explained by the bulk susceptibilities $\left(\bar{\chi}^C, \Delta\bar{\chi}\right)$. Based on the herein proposed model, $f_R$ should directly correspond to the sample averaged mesoscopic frequency contribution $\bar{\Omega}^{Meso}$, assuming chemical exchange and other frequency inducing effects can be neglected. By plugging the previously mentioned susceptibility values into our WM model, Eq. (28) assuming an axially symmetric fODF, we can attempt to predict the values of A and b describing $f_R$. Making the naive assumption that no water resides in bilayers or intra-axonal space ($\lambda = 0$), and that axons are parallel ($p_2 = 1$), $a' = -12.6\,\text{Hz}$ and $b' = 8.7\,\text{Hz}$. Both are around a factor of 2 higher than reported by Wharton and Bowtell. This is not unexpected since orientation dispersion has not been accounted for so far, and since the mesoscopic contribution depends differently on magnetic susceptibility of spherical inclusions, in comparison to the macroscopic contribution. To get a rough idea of the size of such corrections, assume the orientation dispersion can be described by a Watson fODF. Assuming that porcine optic nerve dispersion is similar to WM in the spinal cord, i.e. around 20 degrees [68], this would result in $p_2 = 0.8$, approximately. Based on human WM[69], we estimate $\lambda \sim 1.6$ with the net result that $a' = -9.4\,\text{Hz}$ and $b' = 7.2\,\text{Hz}$. While this is closer to the measured residual frequency, the estimate still does not agree completely with the experiment. The remaining discrepancy could however be attributed to a negative difference between the susceptibility of the agarose and the fluid inside the nerve. For example, if fluid inside the nerve is slightly more diamagnetic, the fitted $\chi_I$ overestimates the WM susceptibility. If so, the coefficients decrease and come into agreement with $f_R$. While the latter correction is somewhat farfetched, it nevertheless demonstrates that by including dispersion, intra-axonal water and a slightly more diamagnetic tissue fluid compared to agarose, the WM model offers an interpretation of the measured residual frequency. This would however require additional experiments to be established. Another potential reason for the discrepancy could originate from MW which has been found to exhibit large frequency shifts that go beyond the presented WM model[11,13,53,59,78]. Since the experiment by Wharton & Bowtell used a 7T scanner with echo times comparable to $T_2^*$ of MW[57,84], $f_R$ would also be reflect the frequency shift from within the MW. A future aim is therefore to repeat the experiment and include measurements of orientation dispersion with dMRI, acquire multiple echo times to investigate time dependence of the Larmor frequency, examine the effect from agarose chemical shifts and perhaps improve upon the susceptibility matching to eliminate potential non-microstructural corrections. This may lead to a better understanding of how well $f_R$ can be described by the modelled effects of WM magnetic microstructure and what goes beyond.

# 6 | Conclusion

We presented an analytical expression for the measured Larmor frequency shift from a biophysical model of WM magnetic microstructure. The model is arguably the most realistic biophysical model for the frequency shift in WM to date, and includes axons with microscopic WM susceptibility anisotropy, intra-compartmental water, and spherical inclusions with scalar susceptibility such as biologically stored iron. Our analytical results for the measured Larmor frequency shift are described in terms of rotation-invariant susceptibility parameters. The structural orientation dependence from axons is described in terms of a fiber orientation distribution which must be determined independently using e.g. dMRI. We validated our analytical results with numerical simulations and demonstrated the feasibility of inverting our model to estimate underlying rotation-invariant susceptibility parameters on a digital brain phantom.

# Acknowledgement

This study is funded by the Independent Research Fund (grants 8020-00158B and 3103-00144B) and Lundbeck BrainComet (grant R310-2018-3455). Fitting codes are available on request.

## Supplementary captions

**Figure S10 – Simulation (b):** Susceptibility fitting maps. Fitting maps for various numbers of sample orientations and a fixed SNR=50 and maximum tilt angle of 90 degrees.

**Figure S11 – Simulation (b):** Susceptibility fitting maps. Fitting maps for various maximum tilt angles and a fixed SNR=50 and 21 sample orientations.

# Supporting Material: The Larmor frequency shift of a white matter magnetic microstructure model with multiple sources

Here we present the derivation of the mesoscopic Larmor frequency shift $\overline{\Omega}(\mathbf{R})$, Equation (1), from our white matter magnetic microstructure model. This entails finding the total Lorentzian tensor $\mathbf{L}$, Equation (3), which consist of three main parts: **(A1)** the induced frequency shift in cylinders from cylinders, **(A2)** the induced frequency in spheres from the spheres, **(A3)** the induced frequency in either cylinders or spheres from the magnetized spheres or cylinders, respectively. All these contributions will be considered in turn.

Integrals are evaluated using the tables in Gradshteyn and Ryzhik[1], validated numerically and reproduced in supplementary material. References to equations are given as GR(X), where X corresponds to the number of the identity in the original tables.

## Supporting simulation for derivation

Each of the analytically derived frequency shifts derived below are presented with a simple computer simulation to demonstrate the results. Figure S1 gives an overview of the microstructure used for the simulation. We constructed a multilayer cylinder in a 3D grid with dimensions $L^3 = 600^3$ grid units. The cylinder consisted of four layers with radii $R_1 = 30$, $R_2 = 50$, $R_3 = 70$, $R_4 = 90$ grid units and cylinder length $L$. Randomly positioned dots (diameter of 1 grid units) were placed in each of the three water compartments, with a total density $\varepsilon$ ranging from approximately 1% to 20%. Using the discrete Fourier transform with periodic boundary conditions, we computed the k-space indicator function for the cylinder, and for each population of spheres. Using Eqs. (3)-(4) we could numerically compute a ground truth Lorentzian tensors $\mathbf{L}$ from combinations of indicator functions and magnetic susceptibility in k-space. These results were compared with the analytical results derived below to validate our results.

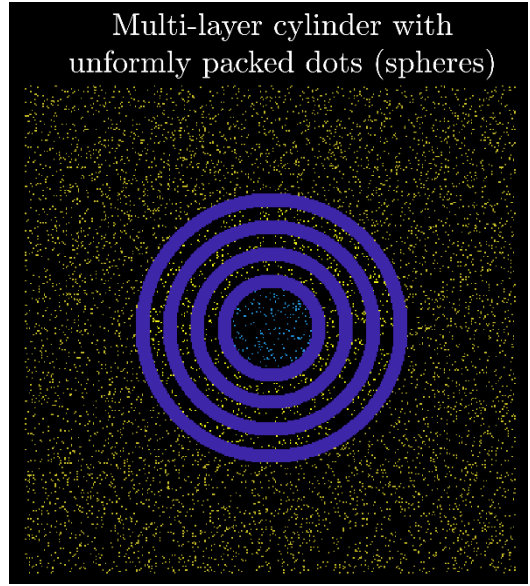

**Figure S1 - Simulation of multi-layer cylinder and spheres:** A 4-layers cylinder are packed in a 3D grid, with uniformly packed dots in all its compartments. Extra-cylindrical dots are shown here in orange, bilayers in purple and intra in cyan. While the dots are uniformly distributed in each compartment, they are still restricted by the presence of the cylinder, which generates a structural anisotropy in their positions.

## S1) Contribution from cylinders with orientation dispersion and susceptibility anisotropy $\Delta\chi$

Here we focus on deriving the Lorentzian tensor $\mathbf{L}$ (cf. Eq. (3)), purely from the population of infinitely long cylinders with arbitrary orientations. Their directions are assumed to be independent of their size and randomly positioned. Each cylinder consists of multiple concentric shells with an associated axially symmetric microscopic susceptibility tensor $\chi^C$ which describes the microscopic content of lipid-protein chains, phospholipid channels etc. as written in Eq. (11).

The Lorentzian tensor $\mathbf{L}$, where we implicitly know that cylinders are both target and source is

$$\mathbf{L} = -\frac{1}{\zeta^W} \int \frac{d\mathbf{k}}{(2\pi)^3} \Upsilon(\mathbf{k}) \Gamma(\mathbf{k}). \tag{S1}$$

Here $\zeta^W$ denotes the total water volume fraction outside of all inclusion types (cylinders of total volume fraction $\zeta^C$ and spheres with a total volume fraction $\zeta^S$). Equation (S1) depends on the tensor-valued cross-correlation tensor $\Gamma(\mathbf{k})$ for cylinders

$$\Gamma(k) = \frac{v^C(k)\chi^C(-k)}{|\mathcal{M}|} - (2\pi)^3 \zeta^C \overline{\chi}^C \delta(k), \text{(cylinder correlation function)}, \tag{S2}$$

where $v^C(k)$ and $\chi^C(k)$ is the structural indicator function and magnetic susceptibility of all the cylinders in k-space, and $\overline{\chi}^C$ is the mesoscopically averaged magnetic susceptibility tensor. The first step in computing $\mathbf{L}$ is to determine $\Gamma(k)$, Eq. (S2), by finding its constituents. We start by re-visiting[2] the k-space indicator function $v^C(k)$ and its mean $\zeta^C$, and then proceed to derive the magnetic susceptibility $\chi^C(k)$ and mesoscopically averaged (bulk) susceptibility $\overline{\chi}^C$.

*Indicator function of cylinders $v^C(k)$*

Positions $r$ within an infinitely long cylinder solid cylinder with radius $R$ displaced $\mathbf{u} = u\hat{\mathbf{u}}$ from the origin can be parametrized by [2]

$$\mathbf{r} = (u + r\cos(\phi))\hat{\mathbf{u}} + r\sin(\phi)\hat{\mathbf{v}} + s\hat{\mathbf{n}}, \tag{S3}$$

where $\hat{\mathbf{n}}$ is a unit vector along the cylinder axis and $\hat{\mathbf{v}}$, $\hat{\mathbf{n}}$ and $\hat{\mathbf{u}}$ are mutually perpendicular. Hence, $(r, \phi, s)$ become local cylinder coordinates (listed here for later convenience). We have previously derived the indicator function for an infinitely long solid cylinder[2,3]. The indicator function $v(k)$ of a single infinitely long solid cylinder is found previously to be

$$\begin{aligned} v(k) &= e^{i\mathbf{k}\cdot\mathbf{u}} \frac{4\pi^2}{k} R J_1(Rk) \delta(\mathbf{k}\cdot\hat{\mathbf{n}}), \quad \text{(infinitely long cylinder)}, \\ &= e^{i\mathbf{k}\cdot\mathbf{u}} v^{2D}(k) 2\pi \delta(\mathbf{k}\cdot\hat{\mathbf{n}}), \end{aligned} \tag{S4}$$

where

$$v^{2D}(k) = \frac{2\pi}{k} R J_1(Rk) \tag{S5}$$

is the 2D indicator function of the cylinder cross-section. $R$ denotes the radii of the cylinder. The volume fraction of the infinite cylinder is understood as $\zeta_1 = 2\pi L R^2 / |\mathcal{M}|$, $L \to \infty$, given as the limit of a cylinder of length 2L going to infinity. The indicator function $v^C(\mathbf{k})$ of all cylinders is then simply the sum $v^C(\mathbf{k}) = \sum_q v_q(\mathbf{k})$ over all individual cylinders labelled by the index q.

*Magnetic susceptibility of a cylinder $\chi^C(\mathbf{k})$*

Consider a lipid forming part of the myelin associated with the cylinder. The lipid is perpendicular to $\hat{\mathbf{n}}$ and form an angle $\phi$ to $\hat{\mathbf{u}}$. The direction of the lipid is $\hat{\mathbf{v}}_1 = \cos(\phi)\hat{\mathbf{u}} + \sin(\phi)\hat{\mathbf{v}}$. Hence, $\hat{\mathbf{v}}_1$ is the eigenvector of $\chi$ associated with parallel susceptibility $\chi^\parallel$ in Eq. (11), whereas the perpendicular vectors to the lipid $\hat{\mathbf{v}}_2 = -\sin(\phi)\hat{\mathbf{u}} + \cos(\phi)\hat{\mathbf{v}}$ and $\hat{\mathbf{n}}$ are the eigenvectors corresponding to $\chi^\perp$. Thus, the susceptibility of a lipid inside a cylinder placed at origo is

$$\chi(\phi) = \chi_\parallel \hat{\mathbf{v}}_1 \hat{\mathbf{v}}_1^T + \chi_\perp \left( \hat{\mathbf{v}}_2 \hat{\mathbf{v}}_2^T + \hat{\mathbf{n}}\hat{\mathbf{n}}^T \right)$$
$$= \left( \chi^C - \frac{1}{3}\Delta\chi \right)\mathbf{I} + \Delta\chi \left( \cos^2(\phi)\hat{\mathbf{u}}\hat{\mathbf{u}}^T + \sin^2(\phi)\hat{\mathbf{v}}\hat{\mathbf{v}}^T + \cos(\phi)\sin(\phi)\left( \hat{\mathbf{v}}\hat{\mathbf{u}}^T + \hat{\mathbf{u}}\hat{\mathbf{v}}^T \right) \right) \text{ (cylinder frame)}$$

(S6)

where we introduced

$$\chi^C \equiv \frac{\chi^\parallel + 2\chi^\perp}{3} = \frac{1}{3}\text{Tr}(\chi) \text{ and } \Delta\chi = \chi^\parallel - \chi^\perp.$$  (S7)

It will prove convenient to rewrite $\chi(\phi)$ as

$$\chi(\phi) = \chi_0 + \text{Re}\{\chi_2 e^{2i\phi}\} = \chi_0 + \frac{1}{2}(\chi_2 e^{2i\phi} + c.c).$$  (S8)

Where c.c. denotes the complex conjugation of the second complex term, and $\chi_0, \chi_2$ are the only non-zero coefficient matrices of $\chi(\phi)$ Fourier series

$$\chi_0 = \chi^C \mathbf{I} + \frac{1}{2}\Delta\chi\left( \frac{1}{3}\mathbf{I} - \hat{\mathbf{n}}\hat{\mathbf{n}}^T \right)$$  (S9)

and

$$\chi_2 = \frac{\Delta\chi}{2}\left( \hat{\mathbf{u}}\hat{\mathbf{u}}^T - \hat{\mathbf{v}}\hat{\mathbf{v}}^T - i\left( \hat{\mathbf{u}}\hat{\mathbf{v}}^T + \hat{\mathbf{v}}\hat{\mathbf{u}}^T \right) \right).$$  (S10)

Having obtained the susceptibility for a single lipid in the myelin sheath, the magnetic susceptibility of the whole myelin sheath $\chi^C(\mathbf{k})$ can be determined by multiplying Eq. (S8) with the cylinder indicator function $v(\mathbf{k})$, Eq. (S4). For a solid cylinder, the susceptibility in k-space becomes

$$\chi^C(\mathbf{k}) = 2\pi\delta(\mathbf{k}\cdot\hat{\mathbf{n}})e^{iu\mathbf{k}\cdot\hat{\mathbf{u}}}\int_0^R d\rho\rho \int_0^{2\pi} d\phi\, e^{i\rho k\cos(\psi-\phi)}\left(\chi_0 + \text{Re}\{\chi_2 e^{2i\phi}\}\right). \tag{S11}$$

Here we used $\hat{\mathbf{k}}\cdot\hat{\mathbf{u}} = \cos(\psi)$, $\hat{\mathbf{k}}\cdot\hat{\mathbf{v}} = \sin(\psi)$. The radial integral of Eq. (S11) is

$$\int_0^R d\rho\rho \int_0^{2\pi} d\phi\, e^{im\phi + ik\rho\cos(\phi-\psi)} = 2\pi \int_{r_q}^{R_q} dr\, r\, e^{im(\psi-\pi/2)} J_m(kr)$$

$$= 2\pi e^{im\psi} \begin{cases} \dfrac{RJ_1(Rk)}{k}, & m = 0 \\[6pt] \dfrac{2 - 2J_0(kR) - kRJ_1(kR)}{k^2}, & m = \pm 2 \end{cases} \tag{S12}$$

$$\equiv 2\pi e^{im\psi} \begin{cases} \xi_0(kR), & m = 0 \\ \xi_2(kR), & m = \pm 2 \end{cases}$$

Here we see that $v^{2D}(\mathbf{k}) = 2\pi\xi_0(Rk)$. Having established all the components of $\chi^C(\mathbf{k})$, the k-space magnetic susceptibility of a solid cylinder becomes

$$\chi^C(\mathbf{k}) = \chi_0 v^C(\mathbf{k}) - 4\pi^2\delta(\mathbf{k}\cdot\hat{\mathbf{n}})\text{Re}\{\chi_2 e^{2i\psi}\}\xi_2(kR), \text{ (solid cylinder)}. \tag{S13}$$

Finding the magnetic susceptibility for a cylinder displaced $\mathbf{u}$, consisting of multiple concentric shells is straightforward, as it follows directly from the shift theorem of the Fourier transform and the superposition principle, respectively, where positions $\rho$ within the q'th layer are denoted by the layer radii $r_q < \rho < R_q$:

$$\chi^C(\mathbf{k}) = 4\pi^2\delta(\mathbf{k}\cdot\hat{\mathbf{n}})e^{i\mathbf{k}\cdot\mathbf{u}}\left\{\chi_0 \sum_q \left(\xi_0(kR_q) - \xi_0(kr_q)\right) - \text{Re}\{\chi_2 e^{2i\psi}\}\sum_q \left(\xi_2(kR_q) - \xi_2(kr_q)\right)\right\}, \text{ (multi-layered cylinder)}. \tag{S14}$$

Similarly, we can sum Eq. (S14) for multiple cylinders with different orientations $\hat{\mathbf{n}}$ and distributions of radii, assumed to be independent of each other. The mean (bulk) susceptibility $\bar{\chi}^C$ of Eq. (S14) within $\mathcal{M}$ is given by its value at $k = 0$

$$\bar{\chi}^C = \frac{1}{|\mathcal{M}|}\chi^C(k=0) = \chi_0 \zeta_1. \tag{S15}$$

*Contribution to Lorentzian tensor from autocorrelation of cylinders*

For simplicity we start by calculating the contribution to the Lorentzian tensor $\mathbf{L}$ from the autocorrelation of two cylinders, describing the mean of the self-generated Larmor frequency shift

$$-\frac{1}{\zeta^W}\int\frac{d\mathbf{k}}{(2\pi)^3}\Upsilon(\mathbf{k})\delta(\mathbf{k}\cdot\hat{\mathbf{n}})\left\{8\pi^2\sum_{q'}\left(\frac{\zeta_{R_{q'}}}{R_{q'}}J_1(kR_{q'})-\frac{\zeta_{r_{q'}}}{r_{q'}}J_1(kr_{q'})\right)\cdot\left(\chi_0\sum_q\left(\xi_0(kR_q)-\xi_0(kr_q)\right)\right.\right.$$
$$\left.-\mathrm{Re}\{\chi_2 e^{2i\psi}\}\sum_q\left(\xi_2(kR_q)-\xi_2(kr_q)\right)\right)$$
$$\left.-4\pi^2\frac{\delta(k)}{k}\zeta_1\bar{\chi}_0\right\} \tag{S16}$$

The volume fraction $\zeta_R$ is of a solid cylinder with radius $R$. Eq. (S16) consists of three parts, where the second part depends on $\psi$. We start by looking at the first and third contribution as they are exactly what was considered in our previous study[2], but here we have the tensor susceptibility $\chi_0$ instead of a scalar. The first and third term are thus

$$\frac{\zeta_1(1-\zeta_1)}{\zeta^W}\frac{1}{2}\left(\hat{\mathbf{n}}\hat{\mathbf{n}}^T-\frac{1}{3}\mathbf{I}\right)\chi_0 = \frac{\zeta_1(1-\zeta_1)}{\zeta^W}\left\{\frac{1}{2}\left(\hat{\mathbf{n}}\hat{\mathbf{n}}^T-\frac{1}{3}\mathbf{I}\right)\chi^C\mathbf{I}-\frac{1}{12}\Delta\chi\left(\hat{\mathbf{n}}\hat{\mathbf{n}}^T+\frac{1}{3}\mathbf{I}\right)\right\}. \tag{S17}$$

Left to solve is the $\psi$-dependent term in Eq. (S16). The angular integration can be carried out by rewriting the dipole kernel into trigonometric functions in the eigenspace of the cylinder

$$\int\frac{d\psi}{\pi}\left(\frac{1}{3}\mathbf{I}-\cos^2(\psi)\hat{\mathbf{u}}\hat{\mathbf{u}}^T-\sin^2(\psi)\hat{\mathbf{v}}\hat{\mathbf{v}}^T-\cos(\psi)\sin(\psi)\left(\hat{\mathbf{v}}\hat{\mathbf{u}}^T+\hat{\mathbf{u}}\hat{\mathbf{v}}^T\right)\right)\mathrm{Re}\{\chi_2 e^{2i\psi}\} = \frac{\Delta\chi}{2}\left(\hat{\mathbf{n}}\hat{\mathbf{n}}^T-\mathbf{I}\right). \tag{S18}$$

What is left to calculate is the radial integral for a single cylinder defined as $\lambda_1$

$$\lambda_1 \equiv -\frac{6}{\zeta^W\zeta^C}\int dk\, k\sum_{q,q'}\left(\frac{\zeta_{R_{q'}}}{R_{q'}}J_1(kR_{q'})-\frac{\zeta_{r_{q'}}}{r_{q'}}J_1(kr_{q'})\right)\left(\xi_2(kR_q)-\xi_2(kr_q)\right). \tag{S19}$$

The parameter $\lambda_1$ describes the layer geometry of the cylinder. We have added -6 to ensure $\lambda_1$ is positive and so its contribution to the Lorentzian tensor is weighted similar to Eq. (S17). Using Eq. (S12) for $\xi_2$ and the integral expressions (which follows from GR(6.533.3) and GR(6.573.(1-2)))

$$\int dk \left(J_1(xk)\right) \frac{2-2J_0(ky)-kyJ_1(ky)}{k^2} = \begin{cases} x\ln\left(\dfrac{y}{x}\right), & y > x \\ 0, & y \leq x \end{cases} \quad \text{(S20)}$$

we obtain

$$\lambda_1 = -\frac{6}{\zeta^W \zeta^C}\left(\sum_{q>q'}\left(\zeta_{R_{q'}} - \zeta_{r_{q'}}\right)\ln\left(\frac{R_q}{r_q}\right) - \sum_q \zeta_{r_q}\ln\left(\frac{R_q}{r_q}\right)\right). \quad \text{(S21)}$$

Collecting all the results we obtain for the auto-correlation contribution to $\mathbf{L}$ from a single cylinder

$$-\frac{\zeta_1(1-\zeta_1)}{\zeta^W}\left\{\frac{1}{2}\left(\hat{\mathbf{n}}\hat{\mathbf{n}}^T - \frac{1}{3}\mathbf{I}\right)\chi^C \mathbf{I} - \frac{1}{12}\Delta\chi\left(\hat{\mathbf{n}}\hat{\mathbf{n}}^T + \frac{1}{3}\mathbf{I}\right)\right\} - \zeta_1 \frac{\Delta\chi}{12}\lambda_1\left(\hat{\mathbf{n}}\hat{\mathbf{n}}^T - \mathbf{I}\right), \text{ (single cylinder).}$$

(S22)

*Contribution to Lorentzian tensor from cross-correlation of cylinders*

Left to consider is the contribution to the Lorentzian tensor, Eq. (S1), from cross-correlations between a pair of different cylinders, i.e., the mean frequency shift inside a cylinder generated by a neighboring cylinder. Fortunately, we have previously shown that such a contribution can be neglected when the magnetic susceptibility of the cylinders is scalar. Second, it is well known from previous studies[4,5] that the functional form of the induced field from a cylinder with a radially symmetric tensor susceptibility anisotropy, as described by Eq. (S6), does not change external to the cylinders. This means that our previous results on considering the contribution from cross-correlations are valid here, as two cylinders are always external to each other. We can therefore neglect an explicit calculation (however, we did calculate it explicitly to verify this).

*Total Lorentzian tensor from cylinders as targets and sources*

Summing over N cylinders, and using $\mathbf{T} = \sum_j \hat{\boldsymbol{n}}\hat{\boldsymbol{n}}^T / N$ is the fODF scatter matrix of the cylinders, and impose the distribution of orientations is independent of their distribution in size, we find the Lorentzian tensor from all the cylinders being targets and sources to be

$$\mathbf{L} = -\frac{\zeta^C(1-\zeta^C)}{\zeta^W}\left\{\frac{1}{2}\left(\mathbf{T}-\frac{1}{3}\mathbf{I}\right)\chi^C\mathbf{I} - \frac{1}{12}\Delta\chi\left(\mathbf{T}+\frac{1}{3}\mathbf{I}\right)\right\} - \zeta^C\frac{\Delta\chi}{12}\lambda^C(\mathbf{T}-\mathbf{I}), \text{ (multiple cylinders)}.$$

(S23)

Here we defined $\zeta^C = N\langle\zeta_1\rangle$ as N times the average cylinder volume fraction $\langle\zeta_1\rangle$, and similar $\lambda^C = N\langle\lambda_1\rangle$.

## S2) Lorentzian tensor from spherical inclusions

Next, we derive the mesoscopic contribution from different populations of solid spherical inclusions. For simplicity, we assume that every cylindrical water compartment, i.e., extra-axonal (E), bi-layers (M) and intra-axonal (A), contains a population of spherical inclusions whose size are independent of their positions, The total indicator function of all spheres in k-space is denoted $v^S(\boldsymbol{k})$, their volume fraction $\zeta^S = \zeta^E + \zeta^M + \zeta^A$, and their total magnetic susceptibility $\chi^S(\boldsymbol{k}) = \chi^A(\boldsymbol{k}) + \chi^M(\boldsymbol{k}) + \chi^E(\boldsymbol{k})$. We assume that each population is uniformly positioned within each compartment. Hence, they are positionally restricted by the cylinders especially when their volume fraction increases. Since $\chi^S$ is a scalar, we only have to consider the demagnetization tensor $\mathbf{N}$

$$\mathbf{N} = \frac{1}{\zeta^W}\int\frac{d\boldsymbol{k}}{(2\pi)^3}\Upsilon(\boldsymbol{k})\Gamma(\boldsymbol{k}),$$

(S24)

which depends on the scalar-valued structure-structure correlation function $\Gamma(\boldsymbol{k})$ of all spheres

$$\Gamma(\boldsymbol{k}) = \frac{v^S(\boldsymbol{k})v^S(-\boldsymbol{k})}{|\mathcal{M}|}.$$

(S25)

For spheres we do not need to care about the singular point of $\Gamma(\boldsymbol{k})$ at $k = 0$, as this term integrates to zero. The generic indicator function $v(\boldsymbol{k})$ for a single sphere of radius $R$ positioned at $u\hat{\boldsymbol{u}}$ from the origin can be found using the following vector $\boldsymbol{r}$ on a spherical surface of radius $r$

$$\boldsymbol{r}(r,\varphi,\vartheta) = \hat{\boldsymbol{u}}u + \hat{\boldsymbol{r}}r + \hat{\boldsymbol{\vartheta}}\vartheta r + r\sin(\theta)\varphi\hat{\boldsymbol{\varphi}}. \tag{S26}$$

Here $(\hat{\boldsymbol{r}}, \hat{\boldsymbol{\vartheta}}, \hat{\boldsymbol{\varphi}})$ is the unit spherical vectors so $(r,\vartheta,\varphi)$ become local coordinates in the sphere. The indicator function $v(\boldsymbol{k})$ becomes

$$v(\boldsymbol{r}) = \int_0^R dr'\, r'^2 \int_0^\pi d\vartheta \sin(\vartheta) \int_0^{2\pi} d\varphi\, \delta\!\left(\boldsymbol{r} - \hat{\boldsymbol{u}}u - \hat{\boldsymbol{r}}r' - \hat{\boldsymbol{\vartheta}}\vartheta r' - r'\sin(\theta)\varphi\hat{\boldsymbol{\varphi}}\right), \tag{S27}$$

and in k-space

$$\begin{aligned}v(\boldsymbol{k}) &= \int d\boldsymbol{k}\, e^{i\boldsymbol{k}\cdot\boldsymbol{r}} \int_0^R dr'\, r'^2 \int_0^\pi d\vartheta \sin(\vartheta) \int_0^{2\pi} d\varphi\, \delta\!\left(\boldsymbol{r} - \boldsymbol{u} - \hat{\boldsymbol{r}}r' - \hat{\boldsymbol{\vartheta}}\vartheta r' - r'\sin(\theta)\varphi\hat{\boldsymbol{\varphi}}\right) \\ &= e^{i\boldsymbol{k}\cdot\boldsymbol{u}} 4\pi R^2 \frac{j_1(kR)}{k}, \text{(single sphere)}.\end{aligned} \tag{S28}$$

Notice that the indicator function for a single sphere only has angular dependence in the exponential describing its positions $\boldsymbol{u}$. This means that the autocorrelation function in k-space $v(\boldsymbol{k})v(-\boldsymbol{k})/|\mathcal{M}|$ of a sphere has no angular dependence, since the exponentials cancel. Therefore, when calculating the contribution to $\mathbf{N}$ from sphere's autocorrelations, the angular integration in Eq. (S28) is exclusively over the dipole kernel, which integrates to zero, i.e.

$$\int d\hat{\boldsymbol{k}}\, \Upsilon(\hat{\boldsymbol{k}}) = 0. \tag{S29}$$

Hence, no frequency shift is associated with autocorrelation from the spheres.

Left to consider is the contribution to $\mathbf{N}$ from cross-correlations $v_1(\boldsymbol{k})v_2(-\boldsymbol{k})/|\mathcal{M}|$ between all possible pairs of spheres in every population. For simplicity, consider the contribution to $\mathbf{N}$ from the cross-correlation between two distinct spheres with radius $R_1$ and $R_2$ separated a distance $\Delta\boldsymbol{u} = \boldsymbol{u}_1 - \boldsymbol{u}_2$

$$\frac{1}{\zeta^W} \frac{2(R_1 R_2)^2}{\pi|\mathcal{M}|} \int d\boldsymbol{k}\, \Upsilon(\boldsymbol{k})\, e^{i\boldsymbol{k}\cdot\Delta\boldsymbol{u}} \frac{j_1(kR_1)\, j_1(kR_2)}{k^2}, \text{(two spheres)}. \tag{S30}$$

As the angular dependence of Eq. (S30) is captured only by the displacement $e^{i\boldsymbol{k}\cdot\Delta\boldsymbol{u}}$, it is convenient to rewrite it in terms of a plane wave expansion[6]

$$e^{i\boldsymbol{k}\cdot\Delta\boldsymbol{u}} = 4\pi \sum_l \sum_{m=-l}^l i^l j_l(k\Delta u) Y_l^{m*}(\Delta\hat{\boldsymbol{u}}) Y_l^m(\hat{\boldsymbol{k}}). \tag{S31}$$

Furthermore, it will be convenient to rewrite the dipole kernel (k-space) and dipole field (real space) in terms of spherical harmonics $Y_2^m$ and STF tensors $\mathcal{Y}_{2m}$ [7]

$$\Upsilon(\mathbf{k}) = -\frac{8\pi}{15} \sum_{m=-2}^{2} \mathcal{Y}_{2m} Y_2^m(\hat{\mathbf{k}}), \text{ (k-space)}, \quad \Upsilon(\mathbf{r}) = \frac{6}{15}\frac{1}{r^3} \sum_{m=-2}^{2} \mathcal{Y}_{2m} Y_2^m(\hat{\mathbf{r}}), \text{ (real space)}. \quad (S32)$$

Using the orthonormality of spherical harmonics, the angular integral of Eq. (S30) is

$$\int d\hat{\mathbf{k}}\, \Upsilon(\mathbf{k}) e^{i\mathbf{k}\cdot\Delta\mathbf{u}} = -\frac{32\pi^2}{15} \sum_{m'=-2}^{2} \mathcal{Y}_{2m'} \sum_l i^l j_l(k\Delta u) \sum_{m=-l}^{l} Y_l^{m*}(\Delta\hat{\mathbf{u}}) \int d\hat{\mathbf{k}}\, Y_2^{m'}(\hat{\mathbf{k}}) Y_l^m(\hat{\mathbf{k}})$$
$$= 4\pi j_2(k\Delta u) \left( \Delta\hat{\mathbf{u}}\Delta\hat{\mathbf{u}}^{\mathrm{T}} - \frac{1}{3}\mathbf{I} \right). \quad (S33)$$

Using GR(6.573.1), the radial integration of Eq. (S30) is

$$\int dk\, j_1(kR_1) j_1(kR_2) j_2(k\Delta u) = \frac{\pi}{6} R_1 R_2 \left( \frac{1}{\Delta u} \right)^3 \quad (S34)$$

Combining Eqs. (S32)-(S34), the contribution to $\mathbf{N}$ from the cross-correlations between two distinct spheres are

$$\mathbf{N} = \frac{V_1 V_2}{\zeta^{\mathrm{W}} |\mathcal{M}|} \Upsilon(\Delta\mathbf{u}), \text{ (two spheres)}, \quad (S35)$$

where $V_1 = 4\pi R_1^3/3$ is the first sphere's volume and $\Upsilon(\Delta\mathbf{u})$ is the dipole field in real space. Eq. (S35) tells us that the mean field induced in a sphere from a neighboring sphere is proportional to the elementary dipole field and scaled by the volumes of the spheres. This is luckily expected since the only non-zero term of the multipole expansion of the field from a sphere is the dipolar term.

The next step is to sum over all these dipolar fields between every sphere in $\mathcal{M}$. However, it is in general highly impractical to sum Eq. (S35), whenever spheres are not Poissonian distributed in the whole mesoscopic region $\mathcal{M}$. To proceed, assumptions must be made about the restrictions in positions of spheres in correspondence with the desired model picture. If the spheres are uniformly positioned in a given water compartment, we do not need to account for every position of the spheres, so long as the spheres are smaller than the size of the water compartments - only the shape of the compartment in which they reside. This allows us to smooth the positions of the spherical inclusions (similar to what is done in QSM on the macroscopic scale). We therefore represent a discrete sum of spheres in a given water compartment by an indicator function $\tilde{v}(r)$ for the whole occupying space

$$v(r) \to \tilde{v}(r)\varepsilon, \text{ (Coarse grained indicator function)}, \tag{S36}$$

where $\varepsilon = \zeta/\tilde{\zeta}$ is the density of spheres that normalizes the indicator function so upon integration it has the original volume fraction $\zeta$ of the spheres, instead of the volume fraction $\tilde{\zeta}$ occupied by the coarse-grained indicator function $\tilde{v}(r)$. For example, the sum over all intra-axonal spheres are represented by the indicator function describing the whole intra-axonal compartment and then scaled by the density of intra-axonal spheres. In fact, this approximation is exactly what we employed when defining the indicator function of the lipid chains forming the cylinders. There, the density factor was just absorbed into the microscopic susceptibility tensor $\chi^C$, Eq. (11), for the cylinder layers. Hence, when considering the contribution to $\mathbf{N}$ from the cross-correlations between two populations of spheres, we make the approximation

$$\begin{aligned}\mathbf{N} &= \sum_{qq'} \frac{\varepsilon_q \varepsilon_{q'}}{\zeta^W} \int \frac{d\mathbf{k}}{(2\pi)^3} \Upsilon(\mathbf{k}) \frac{v_q(\mathbf{k})v_{q'}(-\mathbf{k})}{|\mathcal{M}|} \\ &\simeq \frac{\varepsilon_1 \varepsilon_2}{\zeta^W} \int \frac{d\mathbf{k}}{(2\pi)^3} \Upsilon(\mathbf{k}) \frac{\tilde{v}_1(\mathbf{k})\tilde{v}_2(-\mathbf{k})}{|\mathcal{M}|}\end{aligned}, \text{ (two populations of spheres)}. \tag{S37}$$

Here $\tilde{v}_1(\mathbf{k})$ is understood as the coarse-grained indicator function of the first population with density $\varepsilon_1$ etc. When $\tilde{v}_1(\mathbf{k}) = \tilde{v}_2(\mathbf{k})$, we can think of Eq. (S37) as an autocorrelation between the two populations of spheres (not to be confused with autocorrelation between distinct spheres). All the coarse-grained compartments can be written in terms of cylindrical indicator functions, which makes it easy to calculate $\mathbf{N}$, Eq. (S37). For example, the coarse-grained indicator function of extra-axonal spheres can be written as $\tilde{v}(r) = 1 - v^{SC}(r)$, i.e. determined by the indicator function of all the cylinders (here assumed to be solid and thus indicated by $v^{SC}(r)$ with volume fraction $\zeta^{SC}$ to distinguish from hollowed cylinders).

The demagnetization tensors describing the cross contributions between the same population of spheres are

$$\mathbf{N} = \frac{\zeta_1^2}{\zeta^W} \frac{1-\tilde{\zeta}_1}{\tilde{\zeta}_1} \frac{1}{2}\left(\mathbf{T} - \frac{1}{3}\mathbf{I}\right), \text{ (identical populations)} \tag{S38}$$

Figure S2 shows a good agreement between Eq. (S38) to our simulation (see description in the beginning of the supporting material) for each demagnetization tensor contribution between identical populations of intra-axonal spheres (A), spheres in MW (M) or extra-axonal spheres (E), respectively.

Similarly, the demagnetization tensors from cross contributions between different populations of spheres are

$$\mathbf{N} = -\frac{\zeta_1 \zeta_2}{\zeta^W} \frac{1}{2}\left(\mathbf{T} - \frac{1}{3}\mathbf{I}\right), \text{ (different populations)}. \tag{S39}$$

Figure S3 shows a good agreement between Eq. (S39) to our simulation (see description in the beginning of supporting material) for each demagnetization tensor contribution between different populations of intra-axonal spheres (A), spheres in MW (M) or extra-axonal spheres (E), respectively.

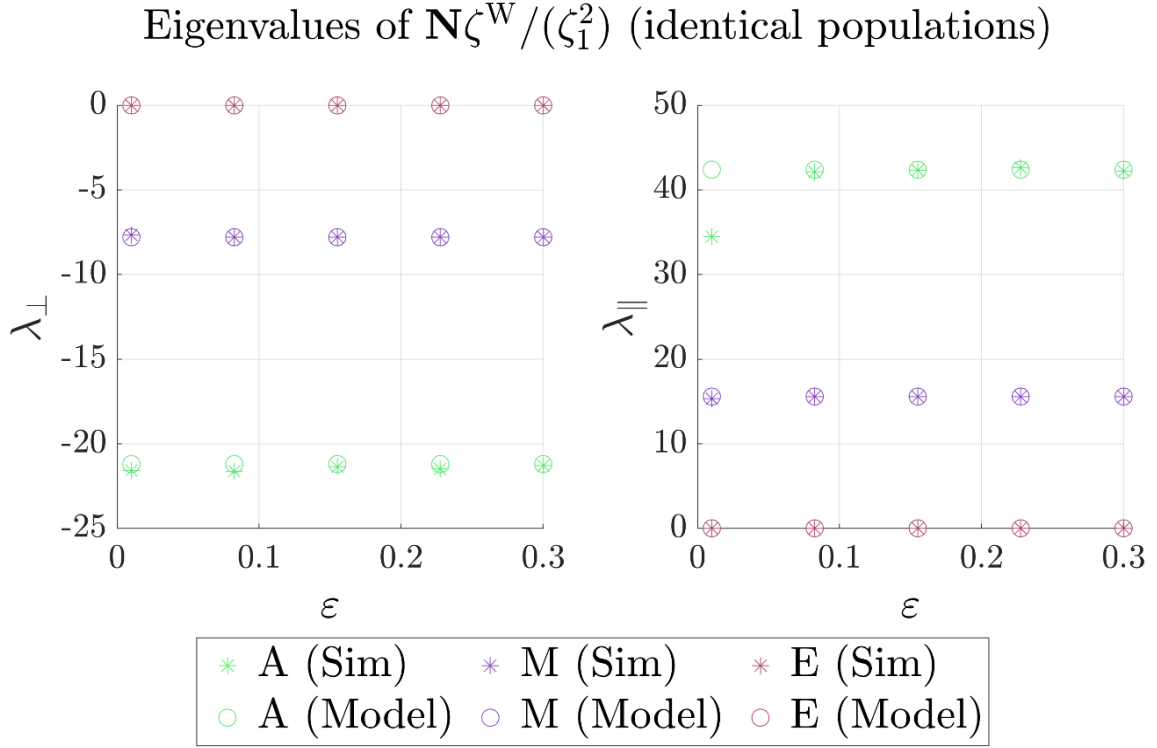

**Figure S2 - Demagnetization tensor eigenvalues from cross-correlations between identical populations of spheres:** Here the target and source are from the same population of spheres. The left figure shows the average perpendicular eigenvalue $\lambda_\perp$, and the right figure shows the parallel $\lambda_\parallel$, wrt. to the axes of the cylinder. The x-axis shows the density $\varepsilon$ of spheres in a given water compartment.

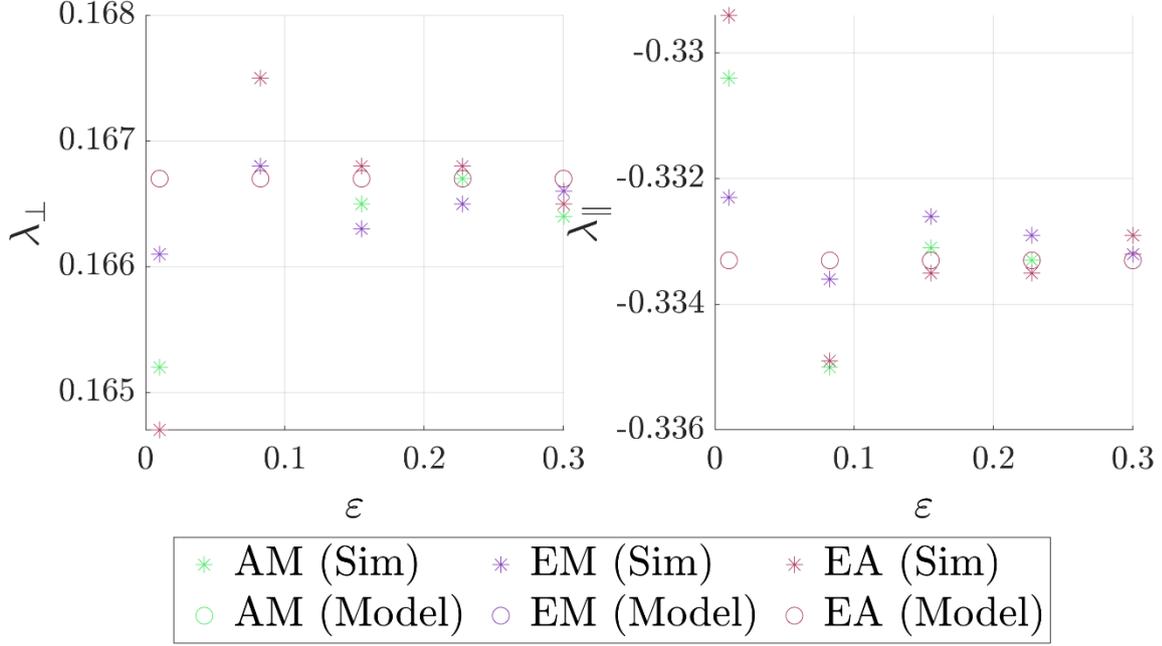

**Figure S3 - Demagnetization tensor eigenvalues from cross-correlations between different populations of spheres:** Here the target and source are different populations of spheres. The left figure shows the average perpendicular eigenvalue $\lambda_\perp$, and the right figure shows the parallel $\lambda_\parallel$, wrt. to the axes of the cylinder. The x-axis shows the density $\varepsilon$ of spheres in a given water compartment.

## S3) Lorentzian tensor from cross-correlations between spherical and cylindrical inclusions

In this section we derive the contribution from cross-correlations between cylindrical and spherical inclusions. Since the cylinders have susceptibility anisotropy, the cross-correlation differs from the induced field from spheres and averaged within the cylinders. We therefore start by considering the demagnetization tensor $\mathbf{N}$ for when spherical inclusions are sources and the cylinders are targets, and then vice versa.

## Demagnetization tensor from spherical sources and cylindrical targets

Consider the demagnetization tensor from a single cylinder layer with radii $r_1$, $R_1$ and a sphere of radius $R_2$ placed at the origin. The cylinder is considered as the target and the sphere generates the field. This demagnetization tensor is described by the scalar-valued cross-correlation function $\Gamma(\boldsymbol{k})$

$$\Gamma(\boldsymbol{k}) = e^{iku\cos(\psi)} \frac{12\pi^2}{k^2} \left( \xi_0(R_1 k) - \xi_0(r_1 k) \right) \zeta_2 \frac{j_1(kR_2)}{R_2} \delta(\boldsymbol{k}\cdot\hat{\boldsymbol{n}}) \\ - 4\pi^2 \zeta_1 \zeta_2 \frac{\delta(\boldsymbol{k}\cdot\hat{\boldsymbol{n}})\delta(k)}{k} \text{, (cylinder target, sphere source),}$$

(S40)

where $\boldsymbol{k}\cdot\boldsymbol{u} = ku\cos(\psi)$. The corresponding contribution to the demagnetization tensor is

$$3\frac{\zeta_2}{\zeta^W R_2} \int dk \int \frac{d\hat{\boldsymbol{k}}}{2\pi} \Bigg\{ \frac{1}{k} \left( \xi_0(R_1 k) - \xi_0(r_1 k) \right) j_1(kR_2) \Upsilon(\hat{\boldsymbol{k}}) \delta(\boldsymbol{k}\cdot\hat{\boldsymbol{n}}) e^{iku\cos(\psi)} \\ - 4\pi^2 \zeta_1 \zeta_2 \frac{\delta(\boldsymbol{k}\cdot\hat{\boldsymbol{n}})\delta(k)}{k} \Bigg\},$$

(cylinder target, sphere source).

(S41)

We start by calculating the first $\psi$-dependent term of Eq. (S41). Rewriting $\Upsilon(\hat{\boldsymbol{k}})$ in terms of trigonometric functions in the basis of the cylinder layer, the angular integral of Eq. (S41) is found to be

$$\int \frac{d\psi}{2\pi} \left( \frac{1}{3}\boldsymbol{I} - \cos^2(\psi)\hat{\boldsymbol{u}}\hat{\boldsymbol{u}}^T - \sin^2(\psi)\hat{\boldsymbol{v}}\hat{\boldsymbol{v}}^T - \cos(\psi)\sin(\psi)\left(\hat{\boldsymbol{v}}\hat{\boldsymbol{u}}^T + \hat{\boldsymbol{u}}\hat{\boldsymbol{v}}^T\right) \right) e^{iku\cos(\psi)} \\ = \left( \frac{1}{3}\boldsymbol{I} - \hat{\boldsymbol{u}}\hat{\boldsymbol{u}}^T \right) J_0(ku) + \left( \hat{\boldsymbol{u}}\hat{\boldsymbol{u}}^T - \hat{\boldsymbol{v}}\hat{\boldsymbol{v}}^T \right) J_1(ku) \frac{1}{ku}.$$

(S42)

The radial integral left to solve is

$$\int dk \frac{1}{k} \left( \xi_0(R_1 k) - \xi_0(r_1 k) \right) j_1(kR_2) \left( \left( \frac{1}{3}\boldsymbol{I} - \hat{\boldsymbol{u}}\hat{\boldsymbol{u}}^T \right) J_0(ku) + \left( \hat{\boldsymbol{u}}\hat{\boldsymbol{u}}^T - \hat{\boldsymbol{v}}\hat{\boldsymbol{v}}^T \right) J_1(ku) \frac{1}{ku} \right). \quad (S43)$$

To proceed, we must consider the integrals in two separate cases: where spheres are either inside the intra-axonal space (i.e. $u + R_2 < r_1$, $R_2 < r_1$) or outside the cylinder in the extra-axonal space (i.e. $R_2 + R_1 < u$).

*Sphere inside cylinder $u + R_2 < r_1$ and $R_2 < r_1$*

When the sphere resides inside the cylinder layer, the two types of radial integrals in Eq. (S43) adds to zero (which follows from GR(6.573.2)), i.e.

$$\sqrt{\frac{\pi}{2R_2}} \int dk \frac{1}{k^{\frac{3}{2}}} J_0(ku) J_{\frac{3}{2}}(kR_2) \left( \xi_0(R_1 k) - \xi_0(r_1 k) \right) = 0, \tag{S44}$$

and

$$\int dk \frac{1}{k^{\frac{5}{2}}} \left( \xi_0(R_1 k) - \xi_0(r_1 k) \right) J_{\frac{3}{2}}(kR_2) J_1(ku) = 0. \tag{S45}$$

Hence no cross-contribution is associated with spheres inside the cylinder layers. This is sensible as this contribution can be seen as the induced frequency at a point (here a sphere) inside a cylindrical shell, which is known to be 0.

*Spheres outside cylinder $R_2 + R_1 < u$*

For the case where the sphere resides outside the cylinder, only one of the two unique integrals yields zero, i.e.

$$\int dk \frac{1}{k} J_0(ku) j_1(kR_2) \left( \xi_0(R_1 k) - \xi_0(r_1 k) \right) = 0. \tag{S46}$$

The first term in Eq. (S43), for a single layer and sphere is thus (cf. GR6.578.1)

$$\frac{3\zeta_2}{\zeta^W R_2} \int dk \frac{1}{uk^{\frac{5}{2}}} \left( \xi_0(R_1 k) - \xi_0(r_1 k) \right) J_{\frac{3}{2}}(kR_2) J_1(ku) \left( \hat{u}\hat{u}^T - \hat{v}\hat{v}^T \right)$$

$$= \frac{1}{2} \frac{\zeta_2}{\zeta^W} \frac{(R_1)^2 - (r_1)^2}{u^2} \left( \hat{u}\hat{u}^T - \hat{v}\hat{v}^T \right),$$

(S47)

If we rotate the cylinders an angle $\phi$ along $\hat{\boldsymbol{n}}$ (rotation axis), Eq. (S47) becomes

$$\frac{1}{2}\frac{\zeta_2}{\zeta^{\mathrm{W}}}\frac{(R_1)^2-(r_1)^2}{u^2}\left(\cos(2\phi)(\hat{\boldsymbol{u}}\hat{\boldsymbol{u}}^{\mathrm{T}}-\hat{\boldsymbol{v}}\hat{\boldsymbol{v}}^{\mathrm{T}})+\sin(2\phi)(\hat{\boldsymbol{v}}\hat{\boldsymbol{u}}^{\mathrm{T}}+\hat{\boldsymbol{u}}\hat{\boldsymbol{v}}^{\mathrm{T}})\right), \text{(cylinder target, sphere source)}.$$

(S48)

Due to the symmetry of $\Gamma(\boldsymbol{k})$ for one layer and external sphere, Eq. (S48) corresponds to the well-known result for the local *microscopic* field outside a cylinder shell, as expected[4]. If we consider Eq. (S48) and sum over all external spheres for each spherical population and all spheres in both the cylinder bilayers and extra-axonal space assuming they are positioned axially symmetric to the layer, the contribution cancels out.

Looking at the last term of Eq. (S41), and summing over both multiple layers, cylinders and spheres, the only contribution left from the cross-correlation between field inducing spheres and reporting cylinders to the Lorentzian tensor $\mathbf{L}=-\chi\mathbf{N}$ from all spherical sources is

$$\mathbf{L}=\frac{\zeta^{\mathrm{C}}\left(\chi^{\mathrm{M}}\zeta^{\mathrm{M}}+\chi^{\mathrm{A}}\zeta^{\mathrm{A}}+\chi^{\mathrm{E}}\zeta^{\mathrm{E}}\right)}{\zeta^{\mathrm{W}}}\frac{1}{2}\left(\mathbf{T}-\frac{1}{3}\mathbf{I}\right), \text{(cylinder targets and sphere sources)}. \quad (S49)$$

Figure S4 shows a good agreement between Eq. (S49) to our simulation (see description in the beginning of supporting material) for each demagnetization tensor contribution between the cylinder as target and intra-axonal spheres (A), spheres in MW (M) or extra-axonal spheres (E) as sources, respectively.

Typically, $\zeta^{\mathrm{C}}/\zeta^{\mathrm{W}} \sim 0.5$ which means that this contribution can only be neglected if the spheres' bulk susceptibilities are sufficiently low.

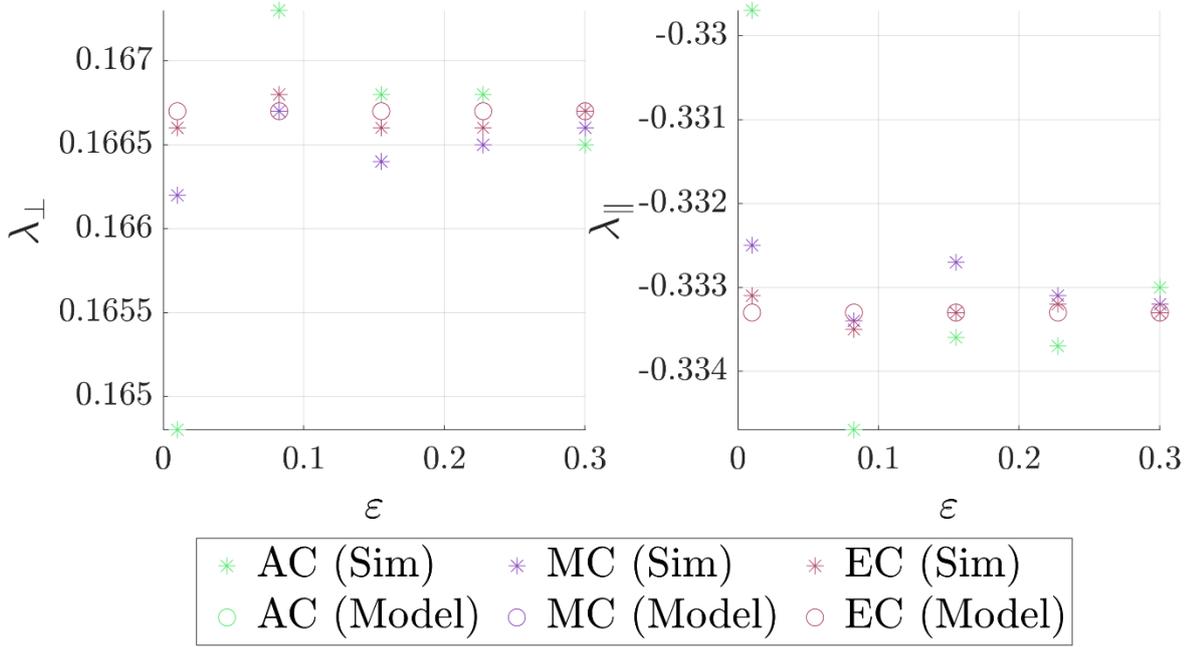

Eigenvalues of $\mathbf{N}\zeta^W/(\zeta^S\zeta^C)$ (cylinder target, sphere sources)

**Figure S4 - Demagnetization tensor eigenvalues from cross-correlation between spheres and cylinders:** Here from a population of spheres induced in the surrounding cylinder layers. The left figure shows the average perpendicular eigenvalue $\lambda_\perp$, and the right figure shows the parallel $\lambda_\parallel$, wrt. to the axes of the cylinder. The x-axis shows the density $\varepsilon$ of spheres in a given water compartment.

Demagnetization tensor from cylindrical sources and spherical targets

We now consider the cross-contribution to the frequency shift inside a reporting sphere of radius $R_1$ and volume fraction $\zeta_1$ placed at the origin generated by a single cylinder layer with radii $r_2, R_2$ and volume fraction $\zeta_2$ positioned at $\boldsymbol{u}$. Its contribution is described by the tensor-valued cross-correlation function $\boldsymbol{\Gamma}(\boldsymbol{k})$ and is given by Eqs. (S14) and (S28)

$$\boldsymbol{\Gamma}(\boldsymbol{k}) = 12\pi^2 \delta(\boldsymbol{k}\cdot\hat{\boldsymbol{n}}) e^{i\boldsymbol{k}\cdot\boldsymbol{u}} \zeta_1 \frac{j_1(kR_1)}{kR_1} \{\chi_0 (\xi_0(kR_2) - \xi_0(kr_2))$$
$$- \mathrm{Re}\{\chi_2 e^{2i\psi}\}(\xi_2(kR_2) - \xi_2(kr_2))\}, \text{ (sphere target and cylinder source)}.$$
$$- 4\pi^2 \zeta_2 \zeta_1 \chi_0 \frac{\delta(k)\delta(\boldsymbol{k}\cdot\hat{\boldsymbol{n}})}{k}$$

(S50)

The first and third term in Eq. (S50) is equivalent to the previously considered cross-contribution between cylinders as the target and spheres the sources (cf. Eq. (S40)). Here we found that only the third contributes to the Lorentzian tensor $\mathbf{L}$ after summing over cylinders and spheres. This means we need only to consider the second and third terms. We start by considering the contribution to the Lorentzian tensor from the second $\psi$-dependent term of $\boldsymbol{\Gamma}(\mathbf{k})$

$$\frac{3}{\pi R_1}\frac{\zeta_1}{\zeta^{\mathrm{w}}}\int dk\int d\hat{\mathbf{k}}\,\Upsilon(\mathbf{k})\delta(\mathbf{k}\cdot\hat{\mathbf{n}})e^{i\mathbf{k}\cdot\mathbf{u}}\operatorname{Re}\{\chi_2 e^{2i\psi}\}\bigl(\xi_2(kR_2)-\xi_2(kr_2)\bigr)j_1(kR_1). \tag{S51}$$

The angular integral of Eq. (S51) is

$$\begin{aligned}&\frac{1}{\pi}\int d\hat{\mathbf{k}}\,\Upsilon(\mathbf{k})\delta(\mathbf{k}\cdot\hat{\mathbf{n}})e^{i\mathbf{k}\cdot\mathbf{u}}\operatorname{Re}\{\chi_2 e^{2i\psi}\}\\ &=\frac{\Delta\chi}{2}\left((\hat{\mathbf{n}}\hat{\mathbf{n}}^{\mathrm{T}}-\mathbf{I})J_0(ku)-\frac{1}{6}(\hat{\mathbf{u}}\hat{\mathbf{u}}^{\mathrm{T}}-\hat{\mathbf{v}}\hat{\mathbf{v}}^{\mathrm{T}})J_2(ku)\right)\end{aligned} \tag{S52}$$

Plugging Eq. (S52) into Eq. (S51) yields the angular integral (excluding some front factors)

$$\int dk\left((\hat{\mathbf{n}}\hat{\mathbf{n}}^{\mathrm{T}}-\mathbf{I})J_0(ku)-\frac{1}{6}(\hat{\mathbf{u}}\hat{\mathbf{u}}^{\mathrm{T}}-\hat{\mathbf{v}}\hat{\mathbf{v}}^{\mathrm{T}})J_2(ku)\right)\bigl(\xi_2(kR_2)-\xi_2(kr_2)\bigr)j_1(kR_1). \tag{S53}$$

Again, Eq. (86) has to be considered separately, for a sphere inside or outside the cylinder.

*Sphere outside cylinder $R_1 + R_2 < u$*

The first radial integral including $J_0(ku)$ in Eq. (S53) is zero which follows from GR(6.573.1), and the independence of $J_0(kr_2)$ and $J_0(kR_2)$ in the integrand for such combinations of Bessel functions and powers[8,9]

$$\int dk\, J_0(ku)\bigl(\xi_2(kR_2)-\xi_2(kr_2)\bigr)j_1(kR_1)=0, \tag{S54}$$

The last integral in Eq. (S53) relating to $J_2(ku)$ also integrates to zero, which follows from GR(6.573.2) and GR(9.180.4)

$$\frac{1}{R_1}\frac{\zeta_1}{\zeta^{\mathrm{w}}}\frac{1}{2}\int dk\, J_2(k\Delta u)\bigl(\xi_2(kR_2)-\xi_2(kr_2)\bigr)j_1(kR_1)=0. \tag{S55}$$

Hence, Eq. (S53) is zero.

*Sphere inside cylinder* $u + R_1 < r_2$ and $R_1 < r_2$

The last to consider is contributions from spheres residing within the cylinder (i.e. $u + R_1 < r_2$, $R_1 < r_2$). The only non-zero radial integral of Eq. (S53) comes from $J_0(ku)$ (which follows from GR(6.573.1)), which is independent of $J_0(ku)$ [8,9]. Using GR(6.577.2), Eq. (S53) becomes

$$\sqrt{\frac{\pi}{2R_1}} \int dk \, \frac{J_{\frac{3}{2}}(kR_1)}{\sqrt{k}} J_0(ku) \left( \xi_2(kR_2) - \xi_2(kr_2) \right) = -\frac{1}{3} R_1 \ln \frac{r_2}{R_2}. \tag{S56}$$

Here we only considered a cylinder made up of a single layer. This means that we first have to sum over spheres inside a single cylinder, either being in the intra-axonal space or in the MW bilayers. Summing over all intra-axonal spheres and cylinder layers constituting the multi-layered cylinder of the single cylinder we can define a new lambda parameter $\lambda_1^A$

$$\lambda_1^A = \zeta_1^A \frac{6}{\zeta^W \zeta_2} \sum_q \ln\left(\frac{r_q}{R_q}\right), \text{ (intra-axonal spheres)}, \tag{S57}$$

which acts as a correction to the previously derived $\lambda_1$ parameter, Equation (S21), for the cylinders. Here $\zeta_1^A$ is understood as the total volume of spheres in the intra-axonal space of a single cylinder. A similar correction arise from spheres within the MW bilayers, where we use $\zeta_q^M$ to indicate the total volume fraction of the spheres inside the q'th MW layer

$$\lambda_1^M = \frac{6}{\zeta^W \zeta_2} \sum_{q'>q} \zeta_q^M \ln\left(\frac{r_{q'}}{R_{q'}}\right), \text{ (myelin water spheres)}. \tag{S58}$$

Left to do is summing over N cylinders in Eqs. (S57) and (S58). For this we define a combined parameter $\lambda^S \equiv \lambda^A + \lambda^M = N\left(\langle \lambda_1^A \rangle + \langle \lambda_1^M \rangle\right)$. Since $\lambda^A + \lambda^M$ has an opposite sign of $\lambda^C$, the presence of spherical inclusions within the intra-axonal space and MW bilayers effectively reduce the Larmor frequency shift caused by susceptibility anisotropy $\Delta\chi$. This makes sense as it reduces the water fraction reporting the Larmor frequency shift from within that compartment.

Using Eqs. (S49), (S53) and $\lambda^S$, we obtain for the Lorentzian tensor from cylindrical sources and spherical targets

$$\mathbf{L} = -\lambda^S \frac{\Delta\bar{\chi}}{12}(\mathbf{T}-\mathbf{I}) + \frac{\zeta^S}{\zeta^W}\left(\bar{\chi}^C \frac{1}{2}\left(\mathbf{T}-\frac{1}{3}\mathbf{I}\right) - \frac{\Delta\bar{\chi}}{12}\left(\mathbf{T}+\frac{1}{3}\mathbf{I}\right)\right), \quad \text{(S59)}$$

(Cylinder sources, sphere targets).

Figure S5 shows a good agreement between Eq. (S59) to our simulation (see description in the beginning of the supporting material) for each Lorentzian tensor contribution between the cylinder source and intra-axonal spheres (A), spheres in MW (M) or extra-axonal spheres (E) as target, respectively.

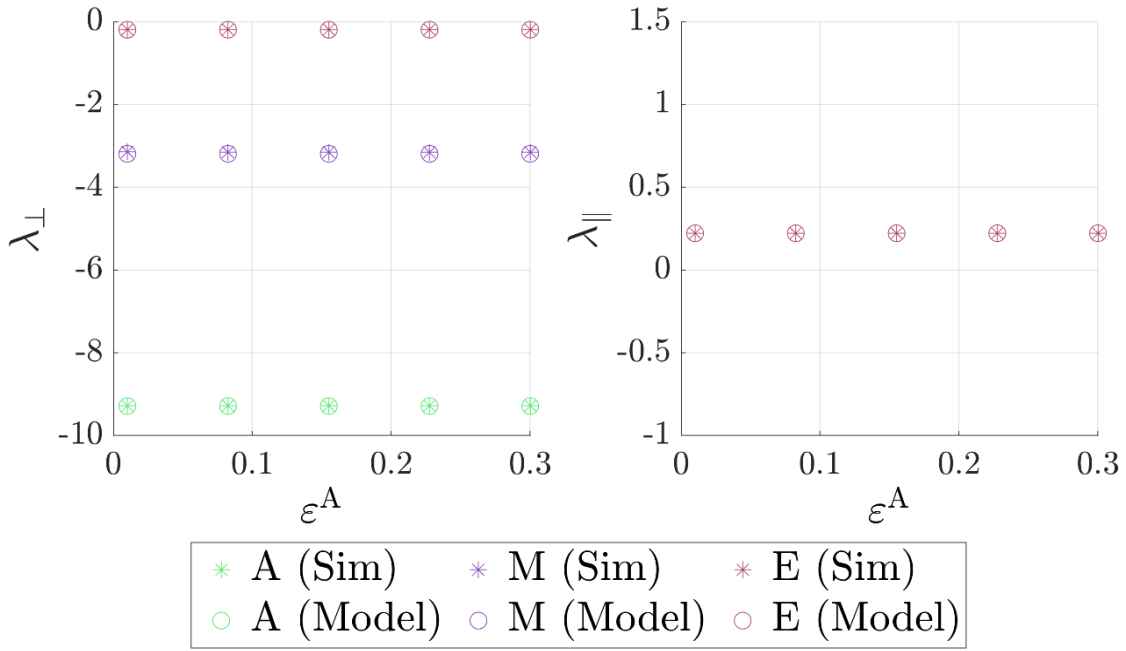

**Figure S5 - Lorentzian tensor eigenvalues from cross-correlation between cylinders and spheres:** Here from cylinders with susceptibility set to unity $\Delta\chi = \chi^C = 1$, induced in a population of spheres. The ground truth signal was made from simulations, where every cylinder layer was associated a magnetic susceptibility tensor, which was computed using $\chi^C(\phi)$, cf. Eq. (S6). The left figure shows the average perpendicular eigenvalue $\lambda_\perp$, and the right figure shows the parallel $\lambda_\parallel$, wrt. to the axes of the cylinder. The x-axis shows the density $\varepsilon$ of spheres in a given water compartment.

## S4) Lorentzian tensor from relaxed myelin water

Last to consider is the cross-contribution to the frequency shift when MW is fully relaxed. If MW is not relaxed, then $\zeta^W = 1 - \zeta$, i.e. equal to the negated volume fraction of the whole magnetized structure. In that case, no additional frequency contributions must be considered. However, when MW is fully relaxed, then $\zeta^W = 1 - \zeta - \zeta^{MW}$, where $\zeta^{MW}$ is the volume fraction of all the water in the MW compartment. In addition, this gives rise to two additional Lorentzian tensors which must be subtracted from our earlier results, since they include the frequency shift in MW: The first describes the frequency shift induced inside MW from cylinders with magnetic susceptibility $\chi^C$, and the second describes the frequency shift induced inside MW from all spheres with total magnetic susceptibility $\chi^S$. Its contribution to the mesoscopic frequency shift is however easy to deduce, at it is basically the same frequency contribution as for the spheres in MW – the only difference is, that upon coarse-graining this MW compartment, the density now relates to the water content $\zeta^{MW}$ and not sphere content $\zeta^M$. In total we get the Lorentzian tensor from MW

$$\mathbf{L} = -\lambda^{MW} \frac{\Delta \bar{\chi}}{12}(\mathbf{T}-\mathbf{I}) + \frac{\zeta^{MW}}{\zeta^W}\left( \bar{\chi}^C \frac{1}{2}\left(\mathbf{T}-\frac{1}{3}\mathbf{I}\right) - \frac{\Delta \bar{\chi}}{12}\left(\mathbf{T}+\frac{1}{3}\mathbf{I}\right)\right)$$
$$- \frac{\bar{\chi}^M}{\zeta^W}\left(1-\zeta^{MW}-\zeta^M-\zeta^A-\zeta^E-\zeta^C\right)\frac{1}{2}\left(\mathbf{T}-\frac{1}{3}\mathbf{I}\right) + \frac{\zeta^{MW}\left(\bar{\chi}^A+\bar{\chi}^E\right)}{\zeta^W}\frac{1}{2}\left(\mathbf{T}-\frac{1}{3}\mathbf{I}\right). \quad (S60)$$

where

$$\lambda^{MW} = N\langle \lambda_1^{MW}\rangle, \quad \lambda_1^{MW} = \frac{6}{\zeta^W \zeta_2}\sum_{q'>q}\left(\zeta_{R_q}^C - \zeta_{r_q}^C - \zeta_q^M\right)\ln\left(\frac{R_{q'}}{r_{q'}}\right). \quad (S61)$$

The single cylinder parameter $\lambda_1^{MW}$ describes the correction for all myelin water within a single cylinder of volume fraction $\zeta_2$.

## S5) Total Mesoscopic frequency shift $\bar{\Omega}^{Meso}$

Combining all contributions to the total mesoscopic frequency shift within our model picture of spherical and cylindrical inclusions, we get the total frequency shift associated with magnetic microstructure within the mesoscopic sphere

$$\bar{\Omega}^{Meso} = \gamma B_0 \hat{\mathbf{B}}^T \mathbf{L} \hat{\mathbf{B}}, \quad (S62)$$

where

$$\mathbf{L} = -\left(\bar{\chi}^C + \bar{\chi}^M\right)\frac{1}{2}\left(\mathbf{T} - \frac{1}{3}\right) + \gamma B_0 \frac{\Delta\bar{\chi}}{12}\left(\mathbf{T}(1-\lambda) + \left(\lambda + \frac{1}{3}\right)\right)$$

$$+ \gamma B_0 \bar{\chi}^A \frac{\left(\zeta^{MW} + \zeta^C - \zeta^A \frac{(1-\tilde{\zeta}^A)}{\tilde{\zeta}^A} + \zeta^E\right)}{\zeta^W} \frac{1}{2}\left(\mathbf{T} - \frac{1}{3}\right)$$

$$+ \gamma B_0 \bar{\chi}^E \frac{\left(\zeta^{MW} + \zeta^C + \zeta^A - \zeta^E \frac{\tilde{\zeta}^C}{1-\tilde{\zeta}^C}\right)}{\zeta^W} \frac{1}{2}\left(\mathbf{T} - \frac{1}{3}\right) \quad . \quad \text{(S63)}$$

When MW is fully relaxed, we get the following effective lambda parameter

$$\lambda = N\langle\lambda_1\rangle = N\left(\langle\lambda_1^C\rangle + \langle\lambda_1^A\rangle + \langle\lambda_1^M\rangle + \langle\lambda_1^{MW}\rangle\right) \quad \text{(S64)}$$

where

$$\lambda_1 = -\zeta_1^{AW} \frac{6}{\zeta^W \zeta_1} \sum_q \ln\left(\frac{r_q}{R_q}\right). \quad \text{(S65)}$$

The volume fraction $\zeta_1^{AW}$ denotes the intra-axonal water fraction inside a single cylinder of volume fraction $\zeta_1$. The largest contribution from the magnetic susceptibilities $\bar{\chi}^E$ and $\bar{\chi}^A$ scale as $\zeta^C/\zeta^W \sim 0.5$, but if the susceptibilities are small, then we may neglect the last two terms of Eq. (S63). When MW is fully relaxed, $\bar{\chi}^{MW}$ contributes to $\bar{\Omega}^{Meso}$ in an equal footing as $\bar{\chi}^C$.

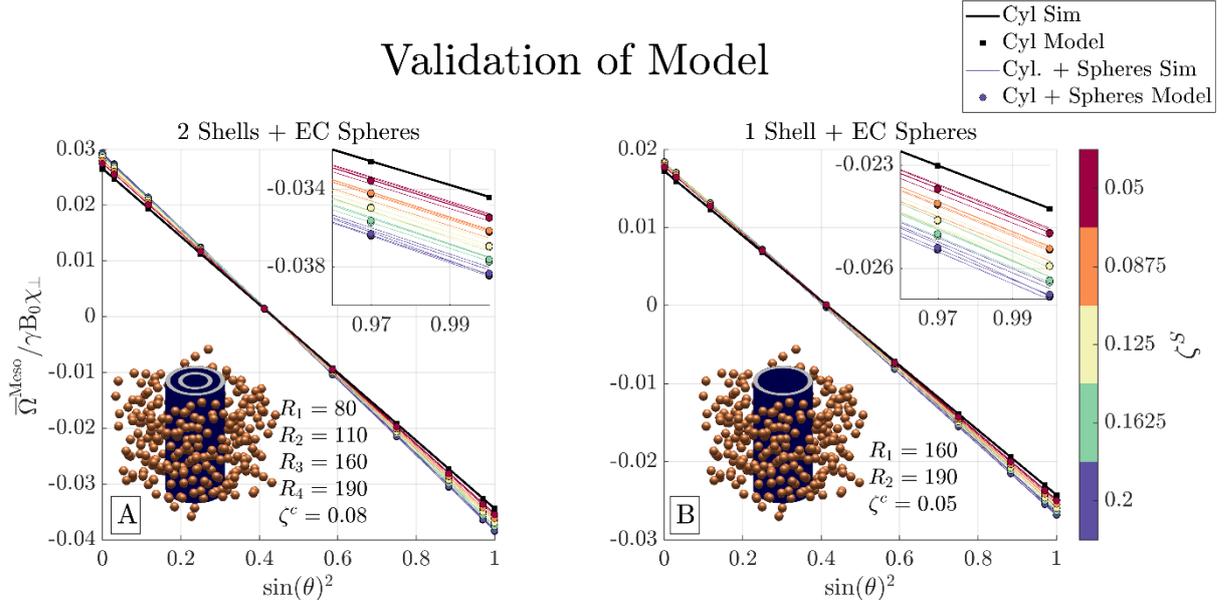

**Figure S6 – Mesoscopic frequency shift $\overline{\Omega}^{\text{Meso}}$ from a multilayer cylinder and external spheres: A** shows $\overline{\Omega}^{\text{Meso}}$ from a 2-layer cylinder with radii and volume fraction shown above. The frequency shift was simulated on $800^3$ grid units. The black line shows $\overline{\Omega}^{\text{Meso}}$ without spheres, and the colored lines indicate different volume fractions $\zeta^C$ of external spheres. The different lines for each color represent different radii of the spheres. **B** shows the same as **A**, but for a single layer cylinder.

Since our result for Eq. (S63) does consider any effects associated with a finite sphere radii, as the distribution of spheres in every water compartment was coarse grained (see A2), we include here an additional validation of Eq. (S63) for different sphere radii. Figure 15 shows $\overline{\Omega}^{\text{Meso}} / \gamma B_0 \overline{\chi}^C$ where $\overline{\chi}^C = \Delta\overline{\chi} = -\overline{\chi}^S < 0$ for a single cylinder of one and two layers, with radii $R_i$ and volume fraction $\zeta^C$ described in Figure S6. The frequency shift was simulated on a grid of dimensions $800^3$, similar to the supporting simulation. The volume fraction $\zeta^S$ of spheres external to the cylinder was varied from 0.01 to 0.2 which is indicated by different colored lines. The black line indicates no spheres. For each volume fraction $\zeta^S$, the sphere radii were varied from 8 to 40 grid units and can be seen as the multiple lines for each color. We find that our model agrees with the simulations, not only for $\overline{\Omega}^{\text{Meso}} / \gamma B_0 \overline{\chi}^C$ from the cylinder alone (black line), but also the overall effect from the spheres. A small variation can be seen for each color from varying the radii, but this uncaptured dependence on sphere radius is small compared to the overall magnitude of $\overline{\Omega}^{\text{Meso}} / \gamma B_0 \overline{\chi}^C$.

## S6) Macroscopic contribution $\bar{\Omega}^{\text{Macro}}(\mathbf{R})$

Last to consider is the macroscopic contribution from nearby voxels in the limit of a slowly varying magnetic microstructure on the macroscale. We previously found[2] this to be

$$\bar{\Omega}^{\text{Macro}}(\mathbf{R}) = \gamma B_0 \hat{\mathbf{B}}^{\text{T}} \sum_{\mathbf{R}'} \bar{\mathbf{Y}}(\mathbf{R} - \mathbf{R}') \bar{\chi}(\mathbf{R}') \hat{\mathbf{B}}. \tag{S66}$$

Using Eq. (S15) for the coarse-grained magnetic susceptibility of a cylinder and summing over them all including the bulk susceptibility across all spheres we get

$$\bar{\Omega}^{\text{Macro}}(\mathbf{R}) = \gamma B_0 \hat{\mathbf{B}}^{\text{T}} \sum_{\mathbf{R}'} \bar{\mathbf{Y}}(\mathbf{R} - \mathbf{R}') \left[ \left( \bar{\chi}^{\text{C}}(\mathbf{R}') + \bar{\chi}^{\text{S}}(\mathbf{R}') \right) \mathbf{I} - \frac{\Delta \bar{\chi}(\mathbf{R}')}{2} \left( \mathbf{T}(\mathbf{R}') - \frac{1}{3} \mathbf{I} \right) \right] \hat{\mathbf{B}}. \tag{S67}$$

## S7) Integrals and identities

Here we list all the non-trivial integrals and identities used from the table by Gradshteyn and Ryzhik [1], seventh edition. Equation numbers corresponds to the table numbers in their book. Identities have also been validated numerically.

$$\int_0^1 dx\, x J_\nu(\alpha x) J_\nu(\beta x) = \frac{\beta J_{\nu-1}(\beta x) J_\nu(\alpha x) - \alpha J_{\nu/1}(\alpha x) J_\nu(\beta x)}{\alpha^2 - \beta^2},\ [\alpha \neq \beta, \nu > -1],\ \text{GR}(5.521.1)$$

$$\int_0^a dx\, J_1(x) = 1 - J_0(a),\ \text{GR}(6.511.7)$$

$$x J_{\nu-1}(x) + x J_{\nu+1}(x) = 2x J_\nu(x),\ \text{GR}(8.471.1)$$

$$\int_0^\infty dx\, \frac{1}{x^2}(1 - J_0(ax)) J_1(bx) = \begin{cases} -\frac{b}{4}\left(1 + 2\ln\left(\frac{a}{b}\right)\right),\ [0 < b < a] \\ -\frac{a^2}{4b},\ [0 < a < b] \end{cases},\ \text{GR}(6.533.3)$$

$$\int\limits_0^{2\pi} d\phi \cos(k\cos(\phi))\cos(n\phi) = 2\pi \cos\left(\frac{n\pi}{2}\right) J_n(k), \quad \text{GR}(3.716.18)$$

$$\int\limits_0^\infty dx\, x^{\nu-M+1} J_\nu(bx) \prod_{i=1}^k J_{\mu_i}(a_i x) = 0, \quad M = \sum_{i=1}^k \mu_i,$$

$$\left[ a_i > 0,\ \sum_{i=1}^k a_i < b < \infty,\ -1 < \operatorname{Re}\nu < \operatorname{Re} M + \frac{1}{2}k - \frac{1}{2} \right], \quad \text{GR}(6.573.1)$$

$$\int\limits_0^\infty dx\, x^{\nu-M-1} J_\nu(bx) \prod_{i=1}^k J_{\mu_i}(a_i x) = 2^{\nu-M-1} b^{-\nu}\, \Gamma(\nu) \prod_{i=1}^k \frac{a_i^{\mu_i}}{\Gamma(\mu_i+1)}, \quad M = \sum_{i=1}^k \mu_i,$$

$$\left[ a_i > 0,\ \sum_{i=1}^k a_i < b < \infty,\ 0 < \operatorname{Re}\nu < \operatorname{Re} M + \frac{1}{2}k + \frac{3}{2} \right], \quad \text{GR}(6.573.2)$$

$$\int\limits_0^\infty dx\, x^{\rho-1} J_\lambda(ax) J_\mu(bx) J_\nu(cx) = 2^{\rho-1} a^\lambda b^\mu c^{-\lambda-\mu-\rho}\, \Gamma\!\left(\frac{\lambda+\mu+\nu+\rho}{2}\right) \frac{1}{\Gamma(1+\lambda)\Gamma(1+\mu)\Gamma\!\left(1-\frac{\lambda+\mu-\nu+\rho}{2}\right)} \cdot$$

$$F_4\!\left(\frac{\lambda+\mu-\nu+\rho}{2}, \frac{\lambda+\mu+\nu+\rho}{2}; 1+\lambda, 1+\mu; \frac{a^2}{c^2}, \frac{b^2}{c^2}\right).$$

$$\left[ a > 0,\ b > 0,\ 0 < a+b < c,\ \operatorname{Re}\{\lambda+\mu+\nu+\rho\} > 0,\ \operatorname{Re}\rho < 5/2 \right], \quad \text{GR}(6.578.1)$$

where $F_4$ is the Appell hypergeometric series of two variables

$$F_4(\alpha,\beta;\gamma,\gamma';x,y) = \sum_{m=0}^\infty \sum_{n=0}^\infty \frac{(\alpha)_{m+n}(\beta)_{m+n}}{(\gamma)_m (\gamma')_n\, m!n!} x^m y^n, \quad \text{GR}(9.180.4)$$

$$(\alpha)_m = \prod_{k=0}^{m-1} (\alpha+k),$$

and $(\alpha)_m$ is the Pochhammer symbol.

## S8) Supplementary figures for simulation b)

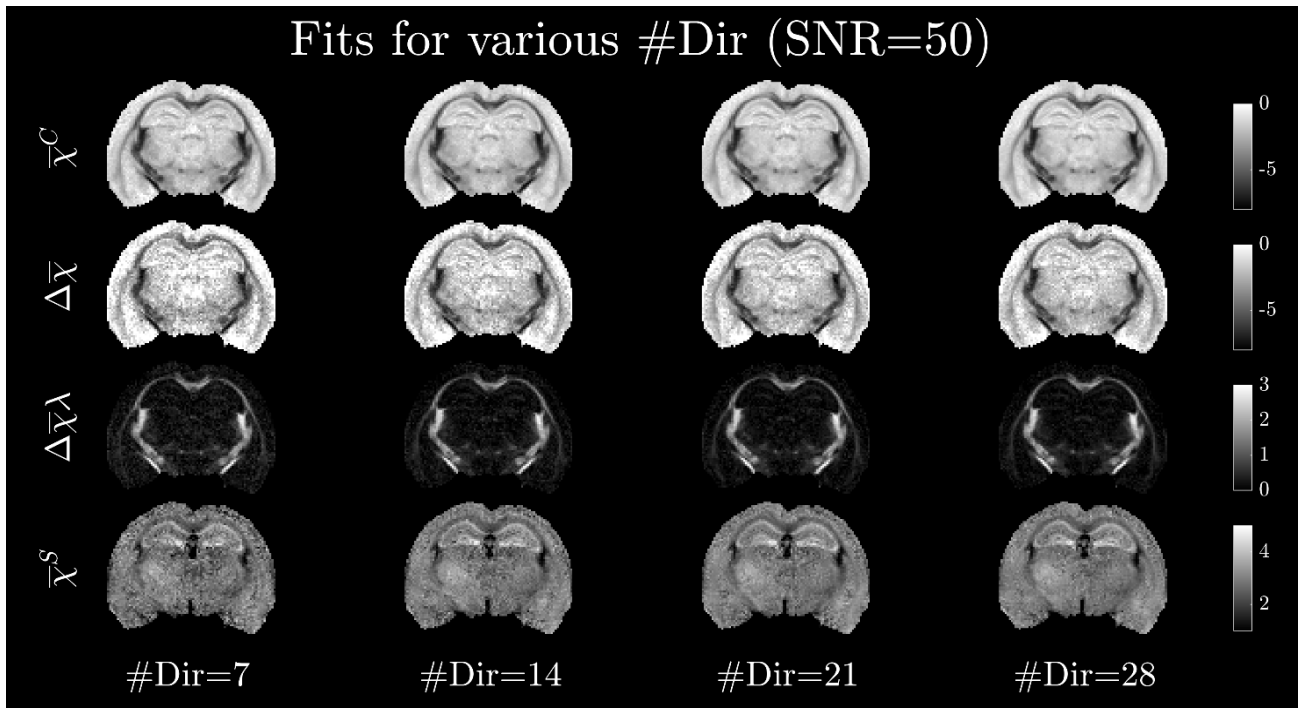

**Figure S7 – Simulation (b):** Susceptibility fitting maps. Fitting maps for various numbers of sample orientations and a fixed SNR=50 and maximum tilt angle of 90 degrees.

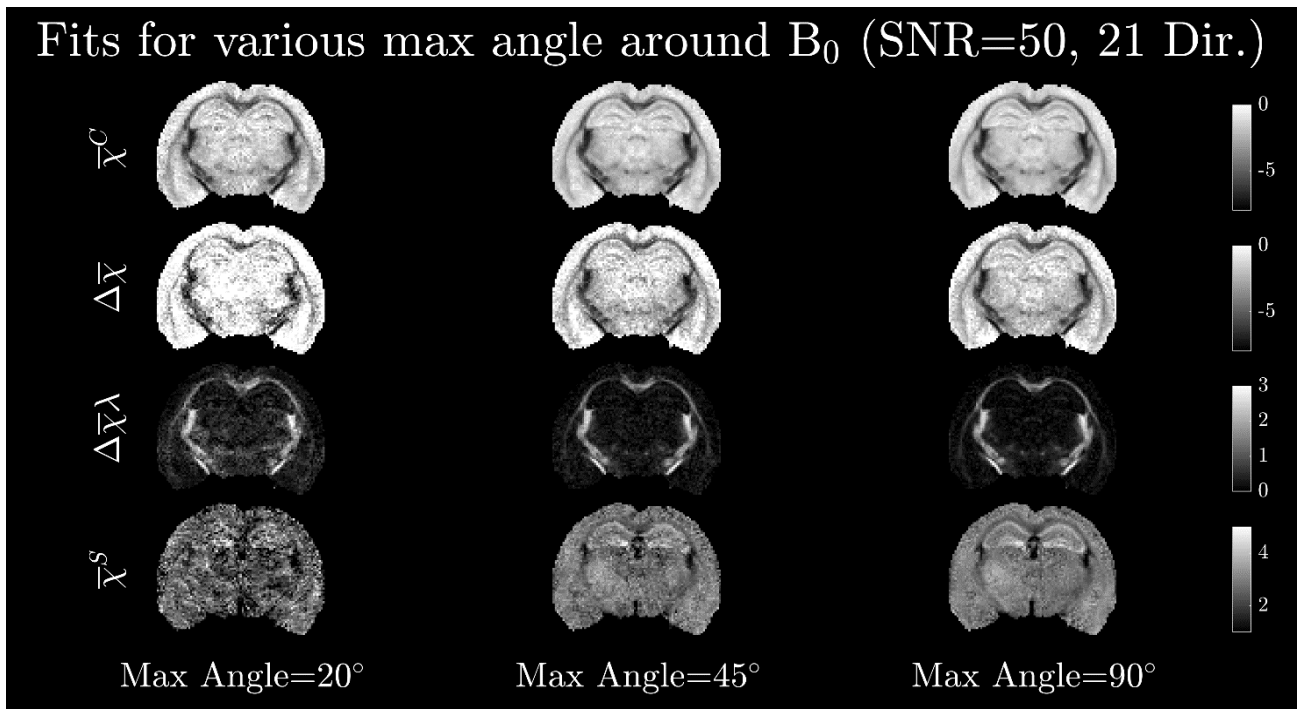

**Figure S8 – Simulation (b):** Susceptibility fitting maps. Fitting maps for various maximum tilt angles and a fixed SNR=50 and 21 sample orientations.